\documentclass[draft]{agujournal2019}
\usepackage{url} 
\usepackage[inline]{trackchanges} 
\usepackage{soul}
\usepackage{miller}
\usepackage{apacite}
\journalname{JGR: Solid Earth}

\begin{document}

%
%


\title{Grain-size effects during semi-brittle flow of calcite rocks}

%
%




\authors{C. Harbord\affil{1}, N. Brantut\affil{1}, D. Wallis\affil{2}}

\affiliation{1}{Department of Earth Sciences, University College London, London, UK}
\affiliation{2}{Department of Earth Sciences, University of Cambridge, Cambridge, UK}




\correspondingauthor{Christopher Harbord}{c.harbord@ucl.ac.uk}



\begin{keypoints}
\item Experimental investigation of semi-brittle flow of calcite rocks with variable grain size
\item Grain size exerts a first-order control on the semi-brittle flow stress of calcite rocks
\item Deformation twinning has limited influence on the strength of calcite rocks
\end{keypoints}

%
%

%
%


\begin{abstract}
We study the role of grain size in the rheological behaviour of calcite aggregates in the semi-brittle regime. We conduct triaxial deformation tests on three rocks, Solnhofen limestone, Carrara marble and Wombeyan marble, with average grain sizes of 5–10~\textmu m, $200$~\textmu m and 2~mm, respectively, at pressures in the range 200–800~MPa and temperatures in the range 20–400$^\circ$C. At all conditions, both strength and hardening rate increase with decreasing grain size. Flow stress scales with the inverse of grain size to a power between 1/3 and 2/3. Hardening rate decreases linearly with the logarithm of grain size. In-situ ultrasonic monitoring reveals that P-wave speed tends to decrease with increasing strain, and that this decrease is more marked at room temperature than at 200 and 400$^\circ$C. The decrease in wave speed is consistent with microcracking, which is more prevalent at low temperature and low pressure. Microstructural observations reveal high twin densities in all deformed samples. Twin density increases with stress, consistent with previous datasets. Spatial distributions of intragranular misorientation indicate that twins are sometimes obstacles to dislocation motion, but this effect is not ubiquitous. Computed slip-transfer statistics indicate that that twins are typically weaker barriers to dislocation glide than grain boundaries, so that their effect on dislocation accumulation and hardening rates is likely smaller than the effect of grain size. Indeed, our data reveal that grain size exerts a first-order control on flow stress and hardening in calcite, whereas twinning may only have a secondary impact on these behaviours.
\end{abstract}

\section*{Plain Language Summary}

Rocks are strongest when their failure mode is mixed between brittle (fracture processes) and crystal plastic (individual grains deform by movement of crystal defects). Unfortunately, we have a limited understanding of how these two mechanisms accommodate deformation together. Yet, both failure modes are sensitive to grain size. We selected three calcite rocks of varying grain size and deformed them at elevated pressure and temperature to simulate coupled brittle and crystal-plastic deformation. Calcite also forms twins (small planar structures in the crystal) when deformed that may act to strengthen the rock during deformation. By monitoring the speed at which sound waves pass through the rock during experiments, we were able to track brittle behaviour. We find that smaller grain size results in greater strength. We observe twins in all deformed samples, but they do not seem to have a big effect on the resulting strength. Microscopic observations of samples and wave-speed measurements show that brittle behaviour is suppressed by increases in pressure and temperature. We suggest a new approach to model these processes that could be used to model coupled brittle and crystal-plastic behaviour on a large scale.

%
%

%


%
%
%
%

\section{Introduction}

In the shallow lithosphere, rocks deform by localised brittle failure and strength is described by a friction law \cite{Scholz2002,Townend2000}. In the lower lithosphere, high temperatures and pressures promote the onset of crystal-plastic deformation mechanisms. Here, strength is described by flow laws sensitive to temperature ($T$) and strain rate ($\dot{\varepsilon}$) \cite{Goetze1972,Evans1995}. At intermediate conditions, deformation is ductile, i.e., remains macroscopically distributed (following the terminology of \citeA{Rutter1986}), and is often termed "semi-brittle", which is characterised by coupled cracking and crystal plasticity. Semi-brittle deformation is likely to support the highest stresses in the lithosphere \cite{Goetze1979,Brace1980}, impacting geodynamic processes \cite{Burov2011} and the maximum depth of earthquake nucleation \cite{Scholz1998}. Despite the significance of semi-brittle flow, there is a paucity of simple models to describe the rheological behaviour of rock in this regime, and the relative importance of key processes, such as friction, tensile cracking, dislocation motion, twinning, or grain-boundary sliding is not well constrained. Thus, to advance our understanding of semi-brittle flow, we must first improve our understanding of interactions among microscale deformation processes.

Calcite rock is an important constituent of tectonic terrains, and undergoes a transition to semi-brittle flow at modest pressure and low temperature \cite<e.g.,>[]{Heard1960,Fredrich1989}. This combination of characteristics has led to numerous laboratory investigations into the rheological behaviour of calcite rock across the brittle-ductile transition (see \citeA{Rybacki2021} and references therein). The main deformation behaviours of calcite rock can be illustrated using the case of Carrara marble, a relatively isotropic, pure calcite aggregate with little initial crack porosity and equant grain shapes with sizes in the range 60–200~$\mu$m. At room temperature, Carrara marble is dilatant and brittle at low pressure ($P<30$~MPa), and its strength is controlled by friction. At intermediate pressures ($30 < P < 300$ MPa), calcite rocks deform in a semi-brittle manner, and strength becomes decreasingly pressure sensitive with increasing pressure \cite{Fredrich1989}. When pressure is high enough ($P>300$ MPa), strength becomes independent of confining pressure and deformation is nondilatant \cite{Edmond1972}. Furthermore, increases in temperature reduce the pressure required to promote semi-brittle flow \cite{Rybacki2021}. \citeA{Edmond1972} and \citeA{Fischer1989} report \textit{in-situ} changes of volumetric strain during deformation of calcite rocks, finding that dilatancy is reduced by increasing pressure. Furthermore, sample volume remains constant when strength is independent of pressure \cite{Edmond1972}. More recently, \citeA{Schubnel2005} and \citeA{Rybacki2021} measured \textit{in-situ} P-wave velocity ($V_\textrm{p}$) and demonstrated that wave-speed decreases during deformation are suppressed by pressure increases. \textit{In-situ} measurements therefore support microstructural observations in implying that increasing pressure suppresses cracking and frictional sliding whilst promoting crystal-plastic processes.

Microstructural observations from calcite rocks deformed in the semi-brittle regime reveal that distributed cracking, twinning, and dislocation motion act together to accommodate strain \cite{Fredrich1989}. \textit{Post-mortem} crack density decreases with increasing confining pressure \cite{Fredrich1989}, approaching zero when strength is independent of confining pressure. Additionally, deformation twins are present in calcite rocks deformed at conditions below 800$^\circ$C \cite{Rybacki2021}, and twin spacing depends on stress \cite{Rowe1990}. Detailed strain measurements at the grain scale reveal that twinning occurs readily in well oriented grains (those with high Schmid factor for twinning), and that it is likely associated with a local backstress that causes hardening of twinned grains \cite{Spiers1979}. In addition, microscale strain mapping also indicates the existence of shear localised along grain boundaries at temperatures $<800^\circ$C, potentially identifying grain-boundary sliding as a possible deformation mechanism \cite{Quintanilla-Terminel2017}.

Another common rheological behaviour of semi-brittle flow in calcite rocks is strain hardening, which can persist to high temperatures \cite<$<800^\circ$C, see>[]{Rybacki2021}. At low temperature, strain hardening can arise from microscopic frictional slip across distributed defects, such as grain boundaries \cite<e.g.,>[]{David2020}. In dislocation-mediated deformation regimes, strain hardening occurs due to an increase in dislocation density caused by inefficient recovery mechanisms (\textit{e.g.}, lack of dislocation climb) \cite{Mecking1981} and is sensitive to microstructure. In particular, grain boundaries can act as barriers to dislocation motion, resulting in increases in strength and sometimes increases in hardening rates with decreasing grain size for most metals \cite<grain-size strengthening, see >[]{Cordero2016}, and some geological materials \cite<\textit{e.g.}, olivine,>[]{Hansen2019}. Alternatively, strain hardening in calcite rocks has been proposed to be controlled by twinning \cite{Rybacki2021}. Twins may not contribute significantly to the total strain, but they could indirectly control strength and hardening by providing additional barriers to dislocations. Indeed, TEM observations show that dislocation densities are elevated adjacent to twin boundaries  \cite{Barber1979,Fredrich1989,Rybacki2021}. In the metallurgy literature \cite{DeCooman2018}, the additional hardening provided by twins is commonly known as twinning-induced plasticity, or TWIP. TWIP originates from observations of high-manganese steels, where high hardening rates and ductility occur as a result of the formation of deformation twins \cite{DeCooman2018}. The hardening rates in these metals are attributed to a "dynamic Hall-Petch effect" in which twinning refines the intracrystalline microstructure and thereby reduces the dislocation mean free path. In this case, twin spacing largely controls strength and hardening and thereby limits sensitivity to grain size.

The semi-brittle regime in calcite rock is thus characterised by many interacting deformation mechanisms. One way to quantify the relative contribution of each mechanism to the overall rheological behaviour is to test the impact of independent variables beyond the usual pressure and temperature conditions and imposed strain rate. One such variable is the grain size. At room temperature, yield stress ($\sigma_\textrm{y}$) and the transition pressure to semi-brittle deformation increase with decreasing grain size \cite{Olsson1974,Fredrich1990}. Grain size also impacts the strength of marble tested at elevated temperature in the dislocation-creep regime, which could be consistent with a Hall-Petch effect \cite{Renner2002}. In the high-pressure, moderate-temperature range ($P>200$~MPa, $T<400^\circ$C) where twinning is ubiquitous, models of twinning-induced plasticity suggest that twin density rather than grain size is the main control on strength \cite{Rybacki2021}. To test this hypothesis, here we explore the role of grain size in semi-brittle rheological behaviour and the brittle-plastic transition in calcite rocks for comparison to the role of twin density. 

We performed a series of experiments at a range of pressures ($P \leq$ 800 MPa) and temperatures ($\leq$ 400$^\circ$C) using calcite rocks spanning two orders of magnitude in grain size: Solnhofen limestone (grain size less than $10$~\textmu m), Carrara marble (grain size on the order of $100$~\textmu m), and Wombeyan marble (grain size on the order of $1$~mm). Experiments are supplemented by \textit{in-situ} measurements of axis-parallel P-wave speed and \textit{post-mortem} microstructural investigations to infer deformation mechanisms. We find that grain size has a first-order impact on both strength and hardening rate, and that the influence of twins is only indirect and likely smaller than anticipated from TWIP models.

\section{Methods}

\subsection{Sample materials}
Three calcite rocks were selected for experiments, with grain size, $D$, spanning three orders of magnitude. Solnhofen limestone is a lithographic limestone with a grain size of 5–10 \textmu m \cite{French2022}, composed of $>99.9\%$ calcite, and with an initial porosity (measured with helium pycnometry) of 4\% \cite{Baud2000}. Carrara Marble is a medium-grained marble with equant grains of size 60–220 \textmu m, comprised of $>99.9\%$ calcite with a porosity of $<$ 0.1\%. In the starting material, most grains exhibit at least one twin set. Wombeyan marble is a coarse-grained marble with a grain size of 1–2 mm, comprised of 96 \% calcite. Grains are equant and typically twinned in at least one plane, with an initial twin density of 60 mm$^{-1}$. This material was obtained thanks to Ian Jackson at the Australian National University, and is the same rock that was used by \citeA{Paterson1958}. See Table \ref{table:petrography} for further petrographic details.

\begin{table}\caption{Composition and physical properties of starting materials}\label{table:petrography}
\begin{tabular}{cccc}
\hline
 &	Solnhofen limestone	& Carrara marble & Wombeyan marble 
\\ \hline
Abbreviation & SL & CM & WM \\
Composition	&	$>$99\% CaCO$_3$	&	$>$99\% CaCO$_3$	&	96\% CaCO$_3$\\
            &                       &                       &   2.5\% MgCO$_3$\\
Porosity (\%)       &	4\%	        &	$<$0.5\%	        &	$<$0.5\%	\\
$d$ (mm)		    &	0.005–0.01  &	0.06–0.22		    & 1–2 \\
Initial $V_{\textrm{p}}$ (m s$^{-1}$)	    &	5600	&	5900	&	- \\\hline
\end{tabular}
\end{table}

\begin{figure}
    \centering
    \includegraphics[width=\textwidth]{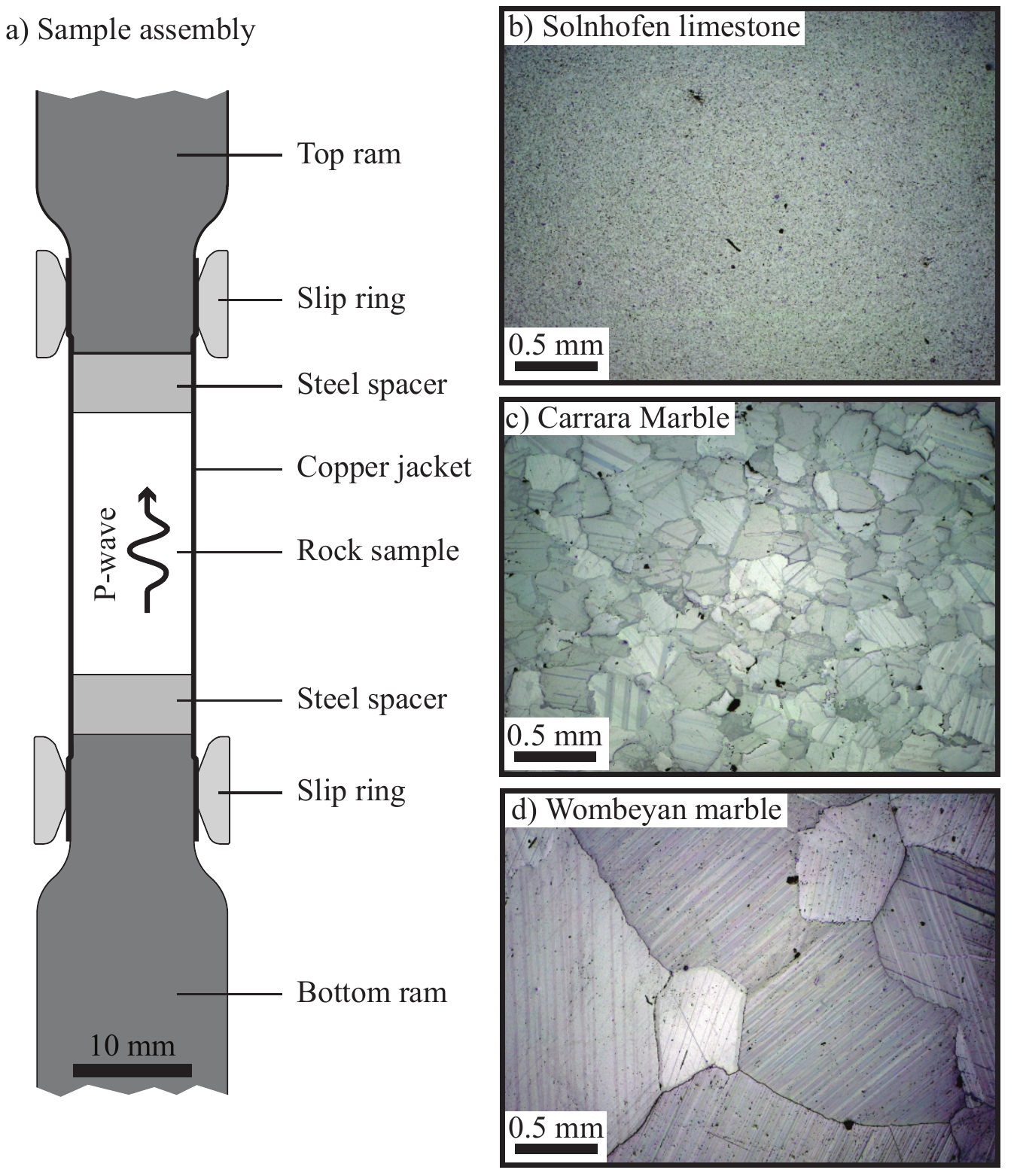}
    \caption{(a) Detail of sample geometry in the pressure vessel.  Insets (b)--(d): Polarised incident-light images of initial sample microstructure of (b) Solnhofen limestone, (c) Carrara marble and (d) Wombeyan marble. }
    \label{fig:Sample_assembly}
\end{figure}

\subsection{Mechanical testing}
A total of 42 experiments were conducted in the "Murrell" gas-medium triaxial apparatus hosted at University College London \cite{Harbord2022}. Rectified cores, 10 mm in diameter and 22 mm in length, were dried in an oven at 70$^\circ$C for at least 24 hours before testing. Samples were inserted into annealed copper jackets of 0.1 mm wall thickness and were swaged onto the deformation rams using a slip ring (Figure \ref{fig:Sample_assembly}a). The jacketed samples were inserted into the pressure vessel and pressurised to the target confining pressure using argon gas. Here the confining pressure is provides the intermediate and minimum principal stresses. Throughout the tests, the measured pressure remained within 5\% of the target pressure. An internal furnace was used to heat samples to $200^\circ$C and $400^\circ$C. Axial load was applied vertically, generating the differential stress, $\sigma$. Axial stress was measured by an external load cell, and axial shortening was measured by a pair of linear variable-displacement transducers. Axial deformation was then applied by a piston that moved at a constant imposed shortening rate of 0.22~\textmu m~s$^{-1}$, equivalent to a strain rate of $10^{-5}$~s$^{-1}$. All samples were deformed to a total strain of 7.5\%. A detailed list of test conditions is given in Table \ref{table:tests}.

\subsection{Data processing}
Several corrections were made to the mechanical data. Seal friction and jacket strength were subtracted from load measurements (as reported in \citeA{Harbord2022}) and displacement due to machine compliance was subtracted from shortening measurements. The differential stress supported by the sample was computed by dividing the corrected force by the cross-sectional area of the sample, which was assumed to linearly increase with deformation consistent with a constant sample volume. Mechanical data were then further processed to obtain estimates of the tangent modulus ($h = \mathrm{\delta}\sigma/\mathrm{\delta}\varepsilon$) by numerical differentiation of the stress data over a moving window of 1\% strain. Uncertainties in seal friction during deformation result in error on the hardening modulus that is $<$ 20 \% of reported values. The yield stress ($\sigma_\mathrm{y}$) is defined as the stress at which $h$ falls to 90\% of the sample-specific Young's modulus determined from elastic loading.

\subsection{\textit{In-situ} P-wave speed measurements}
\textit{In-situ} P-wave speed was measured during tests conducted on Solnhofen limestone and Carrara marble. Measurements were made parallel to the sample axis during deformation using the pulse-transmission method \cite{Birch1961}. Every 10~s during the experiments, a 200~V pulse of $0.4$~\textmu s duration and $2.5$~MHz frequency was sent to a lead-titanate-zirconate ceramic disk mounted centrally in the bottom ram, which was received by a transducer mounted centrally at the top of the sample assembly and recorded digitally at 100~MHz \cite<see >[ for further details]{Harbord2022}. The signal-to-noise ratio was improved by stacking 256 raw traces at each time interval. Changes in axial P-wave speed were computed relative to a reference waveform using cross correlation, and corrected for interfacial delays following the methods outlined in \citeA{Harbord2022}. Measurements of P-wave speed are reported as the change of wave speed ($\Delta V$) and are normalised relative to the wave speed measured at the start of loading ($V_0$). Selected deformation tests were subsequently repeated, and wave-speed changes were found to be reproducible (see Supplementary Material Figure S1).

To gain insight into the microstructural state of samples after deformation during the return to ambient pressure, we continued to perform wave-speed surveys after removal of differential stress and throughout the staged decompression. The change in wave speed during decompression is quantified as the relative change in wave speed ($\Delta V$) normalised by the wave speed measured at the end of deformation ($V_\mathrm{final}$), and when at temperature, after cooling of the sample. These measurements were also complemented by \textit{post-mortem} measurements of wave speed at room pressure (0.1 MPa).

\subsection{Microstructural analysis}

\subsubsection{Optical microscopy}
To investigate deformation microstructures, selected samples were mounted in epoxy, sectioned parallel to the deformation axis, and polished. A set of thin sections was also made for visible-light microscope observations. Imaging using transmitted and reflected light was performed using a Leica DM750P microscope furnished with an ICC50 camera. To quantify the prevalence of intragranular cracks in each sample, we followed the method outlined by \citeA{Fredrich1989}. Samples were imaged using reflected light at $\times4$ magnification, and we counted intersections between cracks (excluding grain boundaries) and a square grid of 2~mm by 2~mm with a spacing of 0.2~mm. These counts were used to determine the resulting crack surface area per volume, $S_\mathrm{v}$ (mm$^2$/mm$^3$).

\subsubsection{Electron microscopy}
Scanning electron microscopy was performed using a Jeol JSM-6480LV scanning electron microscope (SEM) hosted at University College London and a Zeiss Gemini 300 field-emission gun SEM at the University of Cambridge. Electron backscatter diffraction (EBSD) patterns and forescattered electron images were collected using an Oxford Instruments Symmetry detector and AZtec 4.0 acquisition software. Table \ref{table:ebsd} lists the acquisition parameters for each EBSD dataset. All diffraction patterns with acquired with low detector gain. Datasets collected for conventional EBSD were acquired with a reduced number of pixels in the diffraction patterns to increase the mapping speed. Datasets collected for high-angular resolution electron backscatter diffraction (HR-EBSD) were collected with the maximum number of pixels permitted by the detector.

HR-EBSD is a postprocessing technique that analyses distortion of the diffraction patterns to measure lattice rotations and intragranular heterogeneity in elastic strain \cite{Wilkinson2006,Britton2011,Britton2012}. Full details of the technique are given by \citeA{Wallis2019} and here we provide a summary of the key points. One diffraction pattern from the host grain within each mapped area was manually selected to be a reference pattern based on the quality of the diffraction pattern and its position within the map. 100 regions of interest, 256$\times$256 pixels in size, were extracted from all diffraction patterns within the host grain. Each region of interest from each diffraction pattern was cross-correlated with the corresponding region of interest from the reference pattern to determine shifts in their positions. Shifts in the diffraction pattern due to beam scanning were corrected using a calibration determined on an undeformed Si single crystal following \citeA{Wilkinson2006} and the position of the pattern centre was calibrated using diffraction patterns collected over a range of detector insertion distances \cite{Maurice2011}. A deformation-gradient tensor was fit to the field of shifts in each diffraction pattern. The deformation-gradient tensor was decomposed into its symmetric and antisymmetric parts, which respectively give the elastic-strain and rotation tensors \cite{Wilkinson2006}. We used the pattern remapping approach of \citeA{Britton2012}, in which a first pass of cross-correlation measures lattice rotations that are used to rotate each pattern back into the orientation of the reference pattern before a second pass of cross-correlation measures the elastic strain and a small correction to the rotations. The elastic strains were converted to stresses using Hooke's law. The measured stresses are relative to the unknown stress state at the reference point, giving maps of intragranular stress heterogeneity, rather than absolute values. We subtracted the arithmetic mean value of each component of the stress tensor within the map area from each measured value so that the final maps provide stress heterogeneity relative to the unknown mean stress state within each grain \cite{Mikami2015}. Alongside, spatial gradients in the lattice rotations were used to estimate densities of geometrically necessary dislocations (GNDs). Densities of each dislocation type on the slip systems summarised by \citeA{DeBresser1997} were fit to the measurable components of the lattice curvature following the approach applied to quartz by \citeA{Wallis2019}. We emphasise that the stress heterogeneity and GND densities are determined independently, being respectively derived from the distinct symmetric and antisymmetric parts of the deformation-gradient tensor. Data points were filtered out if they had a mean normalised peak height in the cross-correlation function of $<$0.3 or a mean angular error in the fitted deformation gradient tensor of $>$0.004 radians \cite{Britton2011}.

We analysed the probability distributions of the stress heterogeneity to assess whether the stresses are imparted by dislocations. The probability distribution of the stress field of population of dislocations has a characteristic form with tails that depart from a normal distribution towards higher stresses \cite{Jiang2013,Wilkinson2014}. We assess for the presence of these tails using a normal-probability plot, in which the cumulative-probability axis is scaled such that a normal distribution falls on a straight line. Importantly, the tails of the probability distribution, $P(\sigma)$, have a specific form if the stresses, $\sigma$, are imparted by dislocations, whereby $P(\sigma)\rightarrow C\rho |\sigma| ^{-3}$, where $C$ is a constant that depends on the material, type(s) of dislocation, and considered stress component, and $\rho$ is the total dislocation density \cite{Groma1998,Wilkinson2014}. To test whether the measured stress fields exhibit this form, we compute the restricted second moment, $\nu_{2}$, which is a metric that characterises the shape of a probability distribution based on the integral over restricted ranges in stress, calculated as $\nu_{2}(\sigma)=\int_{-\sigma}^{+\sigma} P(\sigma)\sigma^{2}\,d\sigma$ \cite{Wilkinson2014,Kalacska2017}. A plot of $\nu_{2}$ versus ln($\sigma$) becomes a straight line at high stresses if the tails of the probability distribution of the stresses exhibit the form $P(\sigma)\propto|\sigma| ^{-3}$ expected of a population of dislocations \cite{Wilkinson2014,Kalacska2017}. We apply this analysis to the $\sigma_{12}$ component of the stress tensor as this component is the least modified by sectioning the sample and is a shear stress capable of exerting glide forces on dislocations \cite{Wallis2019}. This approach has recently been applied to olivine by \citeA{Wallis2021,Wallis2022}. We include in these plots data from an undeformed Si wafer measured by \citeA{Wallis2022} to provide an indication of the noise level of the stress measurements.

\begin{table}
\centering
\caption{EBSD acquisition parameters for each map} \label{table:ebsd}
\begin{tabular}{lccccc}
\hline
Figure  &   Expt.  &   Lithology    &   Step (\textmu m)  &   Map (X$\times$Y)   &   EBSP pixels (X$\times$Y) \\\hline
Fig. 11a    &   Run0119     &   WM  &   15            &   480$\times$110      &   622$\times$512             \\
Fig. 11b    &   Run0093     &   CM  &   2             &   479$\times$359      &   622$\times$512  \\
Fig. 11c    &   Run0167     &   SL  &   0.15          &   250$\times$191      &   622$\times$512   \\
Fig. 12a    &   Run0093     &   CM  &   2             &   479$\times$359      &   622$\times$512  \\
Fig. 12b    &   Run0093     &   CM  &   0.5           &   450$\times$470      &   622$\times$512  \\
Fig. 12c    &   Run0093     &   CM  &   0.5           &   340$\times$285      &   622$\times$512  \\
Fig. 12d    &   Run0093     &   CM  &   0.5           &   560$\times$560      &   622$\times$512  \\ 
Fig. 13a    &   Run0093     &   CM  &   0.2           &   260$\times$175      &  1244$\times$1024 \\
Fig. 13b    &   Run0093     &   CM  &   0.15          &   300$\times$130      &  1244$\times$1024 \\\hline
\end{tabular}
\end{table}

\subsubsection{Twin density measurements}
Twin density was measured using a combination of forescattered electron images and EBSD maps. For each sample for which twin density was reported, we chose between 20 and 60 grains in a representative forescattered electron image, and measured twin spacing and twin width perpendicular to the selected twin set. Using an EBSD map of the same area, we determined the orientation of each grain and active twin set, and used the angle between the normal to the twin plane and the normal to the section to correct the measured twin width and spacing \cite<the same procedure as used by>[]{Rutter2022}.

\section{Results}
\subsection{Mechanical data and \textit{in-situ} P-wave speed}
\subsubsection{General characteristics}

\begin{table}
\centering
\caption{Table of experiments conducted at a range of conditions. Tests denoted with * are accompanied by measurements of P-wave speed parallel to the specimen axis.} \label{table:tests}
\begin{tabular}{lccccc}
\hline
Experiment	&	Lithology 	& 	$P$ (MPa) & 	$T$ ($^\circ$C)     & $\sigma_5$ (MPa)  & $h_5$ (GPa) \\\hline
Run0147*    &	Solnhofen limestone	&	207		&	20.1     & 466   & 0.34         \\
Run0149*	&	Solnhofen limestone	&	416		&	20.4	 & 535   & 1.57         \\
Run0150*	&	Solnhofen limestone	&	628		&	19.8	 & 517   & 1.85         \\
Run0152*	&	Solnhofen limestone	&	225		&	195	     & 437   & 1.25         \\
Run0153*	&	Solnhofen limestone	&	440		&	180	     & 474   & 2.22         \\
Run0154*	&	Solnhofen limestone &	217		&	393	     & 436   & 1.36         \\
Run0162*	&	Solnhofen limestone &	435		&	394	 & 456  & 1.56         \\
Run0167*	&	Solnhofen limestone	&	610		&	186  & 474  & 2.54         \\
Run0168*	&	Solnhofen limestone	&	617		&	398	 & 416  & 1.84         \\
Run0075	    & 	Carrara marble		& 	200		& 	19.8 & 344  & 1.22         \\
Run0078		& 	Carrara marble		& 	405		& 	19.8   & -    & -	 \\ 	
Run0084		& 	Carrara	marble	    & 	615		& 	19.8   & 386 	& 1.83 \\
Run0086		& 	Carrara	marble  	& 	236		& 	197	 & 187  & 0.73 \\
Run0089		& 	Carrara	marble  	& 	366		& 	204	 & 272  & 1.91 \\
Run0090	    & 	Carrara	marble  	& 	213		& 	390	 & 203  & 0.82 \\
Run0091		& 	Carrara	marble  	& 	378		& 	405  & 186  & 0.72 \\
Run0093		& 	Carrara marble  	& 	599		& 	192	 & 295  & 1.37 \\
Run0094		& 	Carrara	marble  	& 	594		& 	401	 & 198  & 0.65 \\
Run0095		& 	Carrara	marble      & 	602		& 	206  & -    & -    \\
Run0097		& 	Carrara	marble  	& 	597		& 	204	 & -    & -    \\
Run0098		& 	Carrara	marble  	& 	769		& 	20.1 & 377  & 1.67 \\
Run0129*	&	Carrara marble		& 	403		&	17.6 & 385  & 1.67 \\
Run0131*	&	Carrara	marble  	&	202		&	18.1 & 335  & 1.12 \\
Run0132*	&	Carrara	marble      &	407		&	18.2 & 354  & 1.31 \\
Run0137*	&	Carrara	marble  	&	556		&	19.1 & 378  & 1.73 \\
Run0138*    &	Carrara	marble  	&	220		&   172	 & 267  & 1.41 \\
Run0141*	&	Carrara	marble  	&	241		&	387	 & 221  & 1.22 \\
Run0143*	&	Carrara	marble      &	579		&	20.2 & 376  & 1.63 \\
Run0145*	&	Carrara	marble  	&	417		&	370	 & 222  & 0.93 \\
Run0163*	&	Carrara	marble     	&	616		&	171	 & 247  & 1.58 \\
Run0164*	&	Carrara	marble	    &	622		&	368	 & 255  & 1.28 \\
Run0165*	&	Carrara	marble	    &	391		&	195	 & 321  & 2.17 \\
Run0166*	&	Carrara	marble  	&	188		&	183	 & 306  & 1.71 \\
Run0169*	&	Carrara	marble  	&	203		&	16.3 & 341  & 1.36 \\
Run0104		& 	Wombeyan marble 	& 	206		& 	19.3 & 279  & 0.76 \\
Run0106		& 	Wombeyan marble 	& 	594		& 	20.2 & 330  & 1.15 \\
Run0110		& 	Wombeyan marble 	& 	231		& 	207	 & 189  & 1.19 \\
Run0119		& 	Wombeyan marble 	& 	609		& 	187	 & 218  & 0.55 \\
Run0120		& 	Wombeyan marble 	& 	408		& 	19.2 & 338  & 0.75 \\
Run0121		& 	Wombeyan marble 	& 	428		& 	190	 & 226  & 0.71 \\
Run0124		& 	Wombeyan marble 	& 	233		& 	408	 & 128  & 0.08 \\
Run0126		& 	Wombeyan marble 	& 	574		& 	416	 & 100  & --0.02 \\
Run0127		& 	Wombeyan marble 	& 	400	    & 	392	 & -    & -     \\\hline
\end{tabular}
\end{table}

\begin{figure}
\centering
\includegraphics[scale=1]{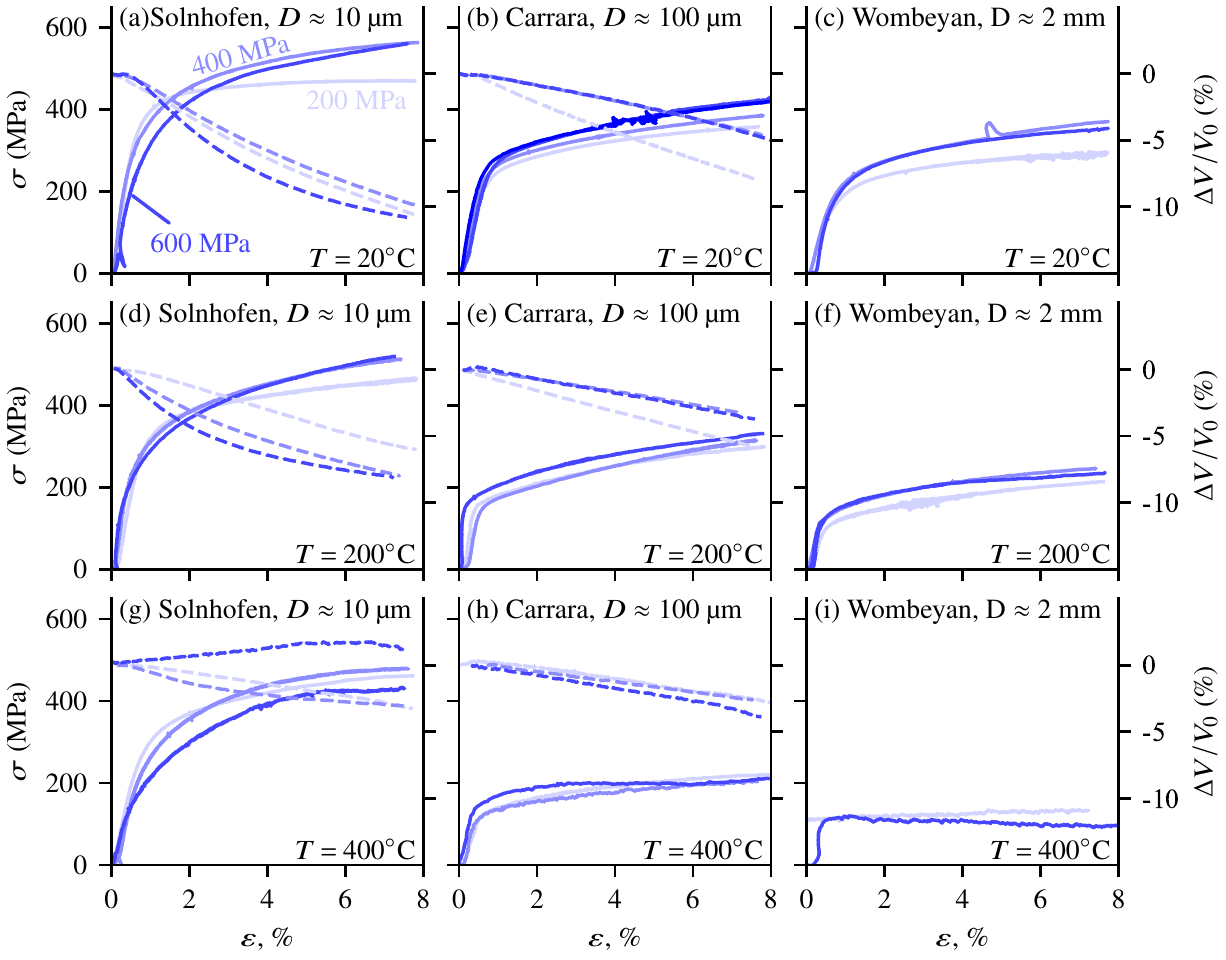}
\caption{Differential stress ($\sigma$, solid lines) and normalised P-wave speed ($\Delta V/V_0$, dashed lines) from constant-strain rate experiments at $P$ = 200, 400, 600, and 800 MPa, indicated by increasing intensity of blue, at 20, 200 and 400$^\circ$C increasing downward in the plot. Plots (a), (d) and (g) correspond to experiments performed on Solnhofen limestone at 20, 200 and 400$^\circ$C respectively. Plots (b), (e) and (h) are tests performed on Carrara marble at 20, 200 and 400$^\circ$C respectively. Plots (c), (f) and (i) are tests performed on Wombeyan marble at 20, 200 and 400$^\circ$C respectively.}
\label{Fig:StrainStressVel}
\end{figure}

The stress-strain behaviour and evolution of \textit{in-situ} P-wave speed are qualitatively similar across the range of conditions and tested materials (Figure \ref{Fig:StrainStressVel}), and are typical of ductile behaviour. Taking the example of Carrara marble at $P = 600$ MPa and $T = 20^\circ$C (Figure \ref{Fig:StrainStressVel}b, solid dark blue curve), the mechanical data is characterised by a rapid linear increase in stress at low strain ($\varepsilon < 0.5$\%), representing elastic loading, during which P-wave speed remains relatively constant (Figure \ref{Fig:StrainStressVel}b, dashed dark blue curve). Above a stress of around 250 MPa, the rate of increase in stress with strain begins to decrease and the P-wave speed begins to decrease concomitantly. Beyond approximately 1.5\% strain, stress continues to increase approximately linearly with strain, which is accompanied by a steady decrease in wave speed. Similar qualitative behaviour occurs in all samples at all conditions tested, with quantitative variations in the yield stress and degree of hardening.

Strain hardening is observed at nearly all test conditions (Figure \ref{Fig:StrainStressVel}), with the exception of Wombeyan marble deformed at $T = 400 ^\circ$C (Figure \ref{Fig:StrainStressVel}i), solid curves). Between $20 ^\circ$C and $200 ^\circ$C, hardening rates are unaffected by temperature increases, however the hardening modulus is lower for all lithologies at $T = 400 ^\circ$C.

\begin{figure}
    \centering
    \includegraphics{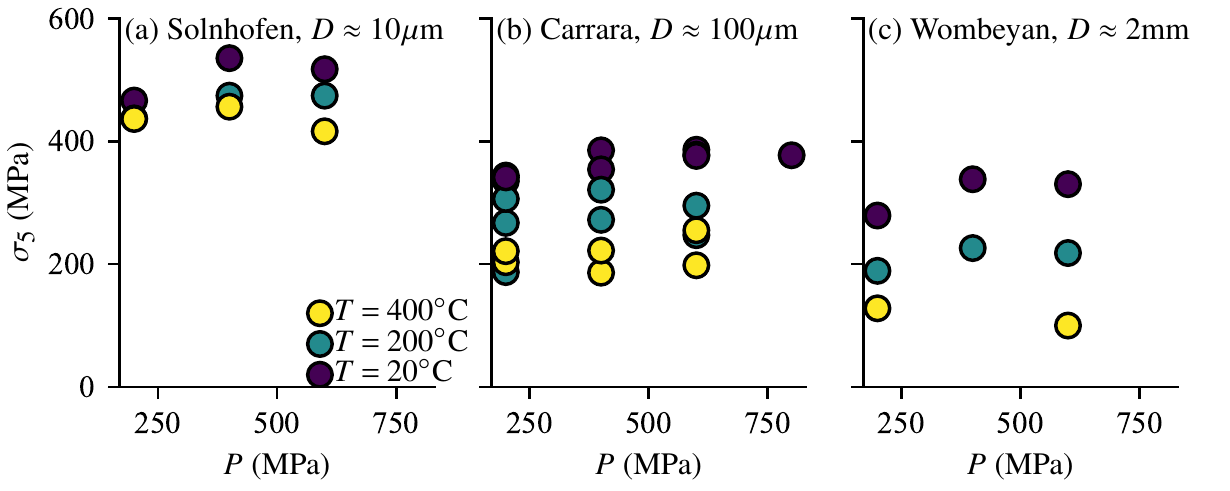}
    \caption{Differential stress at 5\% strain ($\sigma_{5}$) as a function of confining pressure in Solnhofen limestone (a), Carrara marble (b) and Wombeyan marble (c), at temperatures of 20, 200 and 400$^\circ$C. All tests were conducted at a strain rate of $10^{-5}$~s$^{-1}$.}
    \label{fig:pressure_dependency}
\end{figure}

\subsubsection{Effect of pressure and temperature}

\begin{figure}
\centering
\includegraphics[scale=1]{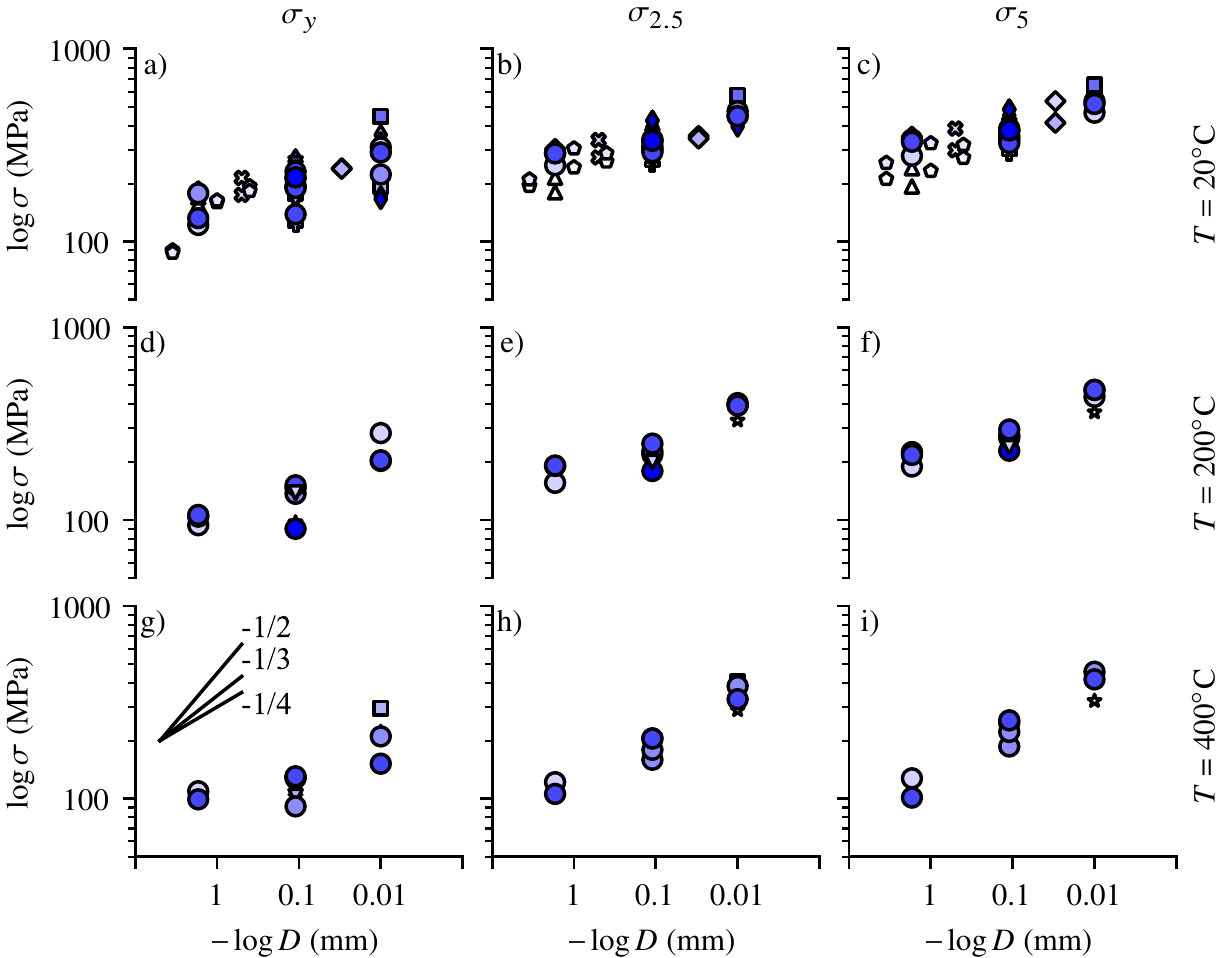}
\caption{Evolution of yield stress ($\sigma_\mathrm{y}$), differential stress at 2.5\% strain ($\sigma_{2.5}$) and 5\% strain ($\sigma_5$) as functions of grain size. Circles correspond to data obtained in this work, and other symbols are taken from the literature (see Table \ref{tab:literature} for list of labels).} 
\label{Fig:D_Strength}
\end{figure}

The absolute strength, \textit{i.e.} the differential stress at a given strain, of all rocks tested decreases with increasing temperature. The temperature sensitivity is greater at elevated pressure (400 and 600~MPa) than at low pressure, and is greater in Wombeyan marble than in Carrara marble and Solnhofen limestone (Figure \ref{fig:pressure_dependency}). 

Both Carrara marble and Wombeyan marble have strengths and yield stresses that tend to increase with increasing pressure (Figure \ref{fig:pressure_dependency}). However, strength seems to become pressure-independent beyond a pressure threshold. In Carrara marble, there are no quantitative differences between the tests conducted at 600~MPa and at 800~MPa, and only small differences can be detected between 400~MPa and 600~MPa. In Wombeyan marble, strength does not change appreciably with pressure above 400~MPa. The pressure sensitivity of strength in Carrara marble and Wombeyan marble is reduced by increasing temperature, and strength becomes pressure independent at a lower pressure as temperature increases. Hardening rates in Carrara marble and Wombeyan marble also increase with increasing pressure, except for at a temperature of $400 ^\circ$C, at which hardening rates are independent of pressure.

Solnhofen limestone exhibits some differences compared to the other lithologies in terms of the pressure dependence of strength. At all conditions, the yield stress of Solnhofen limestone decreases with increasing pressure, in contrast with the other lithologies. In contrast, at 5\% strain, the strength of Solnhofen limestone increases slightly with pressure from 200 to 400~MPa, but decreases at 600~MPa (Figure \ref{fig:pressure_dependency}a).

Wave-speed changes also depend on pressure and temperature. Decreases in axial P-wave speed with strain are smaller at higher temperature in both Solnhofen limestone and Carrara marble. For example, in Carrara marble at a pressure of $200$ MPa and strain of $7.5\%$, the relative wave-speed decrease is 8\% at a temperature of $20^\circ$C (Figure \ref{Fig:StrainStressVel}b ) and is 2.5\% at $400^\circ$C (Figure \ref{Fig:StrainStressVel}h). In Carrara marble, at all temperatures, increasing pressure results in smaller wave-speed decreases during deformation. In contrast, in Solnhofen limestone at temperatures of $20 ^\circ$C and $200 ^\circ$C the final wave-speed drop increases with increasing pressure (Figure \ref{Fig:StrainStressVel}a,d).

\begin{figure}
\centering
\includegraphics[scale=1]{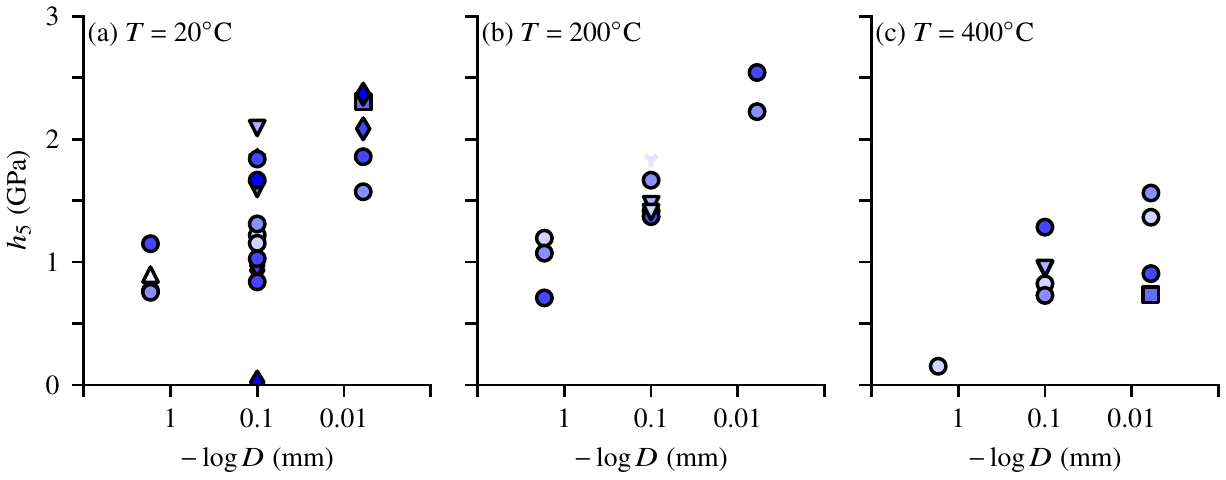}
\caption{Hardening modulus at 5\% strain as a function of grain size, at (a) 20$^\circ$C, (b) 200$^\circ$C and (c) 400$^\circ$C. Circles correspond to data obtained in this work, and other symbols are taken from the literature (see Table \ref{tab:literature} for list of symbols).}
\label{Fig:Hardening}
\end{figure}

\subsubsection{Effect of grain size}

\begin{table}
    \centering
    \caption{Summary of references for literature data presented in Figure \ref{Fig:D_Strength} and \ref{Fig:Hardening}.}
    \label{tab:literature}
    \begin{tabular}{l c c c}
    \hline
        Reference               &  Lithology            &   Grain size              & Legend    \\ \hline      
        \citeA{Paterson1958}    &  Wombeyan marble      &   1–2 mm                 & \includegraphics[scale=.2]{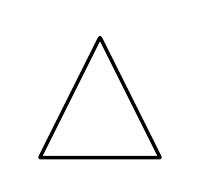} \\
        \citeA{Heard1960}       &  Solnhofen limestone  &   6–10 \textmu m                  & \includegraphics[scale=.2]{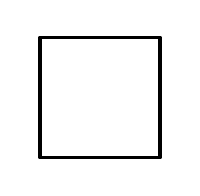} \\
        \citeA{Edmond1972}      &  Solnhofen limestone  &   6–10 \textmu m                   & \includegraphics[scale=.2]{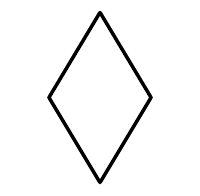}  \\
        \citeA{Edmond1972}      &  Carrara marble       &   60–120 \textmu m                   & \includegraphics[scale=.2]{Figures/LiteratureLegend/EP1972.pdf}  \\
        \citeA{Fredrich1989}    &  Carrara marble       &   60–120 \textmu m                     & \includegraphics[scale=.2]{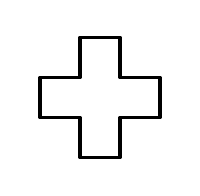}  \\
        \citeA{Rybacki2021}     &  Carrara marble       &   60–120 \textmu m                  & \includegraphics[scale=.2]{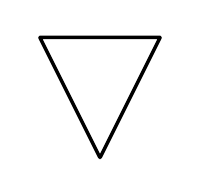}  \\
        \citeA{Fredrich1990}    &  Wombeyan marble      &   1–2 mm                                  & \includegraphics[scale=0.2]{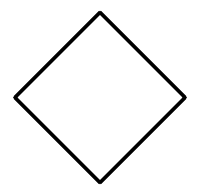} \\
        \citeA{Donath1971a}     &  Beldens marble       &   0.5 mm                                     & \includegraphics[scale=.2]{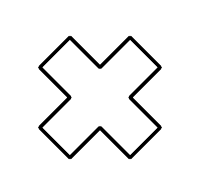}  \\
        \citeA{Mogi1964}        &  Yamaguchi 'Fine marble'   &   0.4 mm                                    & \includegraphics[scale=.2]{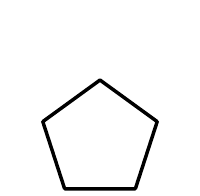}\\
        \citeA{Mogi1964}        &  Mito marble &   1 mm                                      & \includegraphics[scale=.2]{Figures/LiteratureLegend/Mogi1964.pdf}\\
        \citeA{Mogi1964}        &  Yamaguchi 'Coarse marble' &   3.5 mm                                     & \includegraphics[scale=.2]{Figures/LiteratureLegend/Mogi1964.pdf}\\
        \citeA{Rutter1974}      &  Solnhofen limestone  &   6–10 \textmu m                     & \includegraphics[scale=.2]{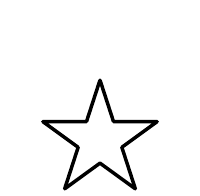}\\\hline
    \end{tabular}
    
\end{table}

Grain-size has a marked effect on the mechanical behaviour. Solnhofen limestone is always the strongest material and Wombeyan marble the weakest, independent of test conditions (Figures \ref{Fig:StrainStressVel} and \ref{Fig:D_Strength}). For example, at a temperature of $20^\circ$C and pressure of 600 MPa, the stress at $7.5\%$ strain is 570 MPa for Solnhofen limestone, 400 MPa for Carrara marble, and 310 MPa for Wombeyan marble (Figure \ref{Fig:D_Strength}).

Grain-size strengthening is further promoted by increases in strain. For example, at a temperature of $20^\circ$C the range of stresses across the grain sizes tested is $\sigma_\mathrm{y} =$ 100–300~MPa at the yield stress, $\sigma_{2.5} =$ 250–500~MPa at 2.5\% strain, and $\sigma_5 =$ 300–600~MPa at 5\% strain (Figure \ref{Fig:D_Strength}). This observation is consistent with measurements of the hardening modulus, which increases with decreasing grain size at all tested temperatures (Figure \ref{Fig:Hardening}).

Wave-speed evolution is also sensitive to grain size, as highlighted by comparing the behaviour of Solnhofen limestone and Carrara marble. The overall decrease in wave speed is nearly always greater for Solnhofen limestone at a given set of conditions than it is for Carrara marble. For example, at $T = 20 ^\circ$C, $P = 600$ MPa and $\varepsilon = 7.5$\%, the wave-speed change is $-5$\% in Carrara marble (Figure \ref{Fig:StrainStressVel}b) but is $-11$\% in Solnhofen limestone (Figure \ref{Fig:StrainStressVel}a). The only exception to this grain-size dependence is at $T = 400^\circ$C and $P = 600$ MPa. At these conditions, the final relative wave-speed change at $\varepsilon = 7.5$\% is $+2$\% in Solnhofen limestone (Figure \ref{Fig:StrainStressVel}g) and -2\% in Carrara marble (Figure \ref{Fig:StrainStressVel} h) dashed dark blue curve).

\subsection{Wave-speed changes during decompression}

\begin{figure}
\centering
\includegraphics[scale=1]{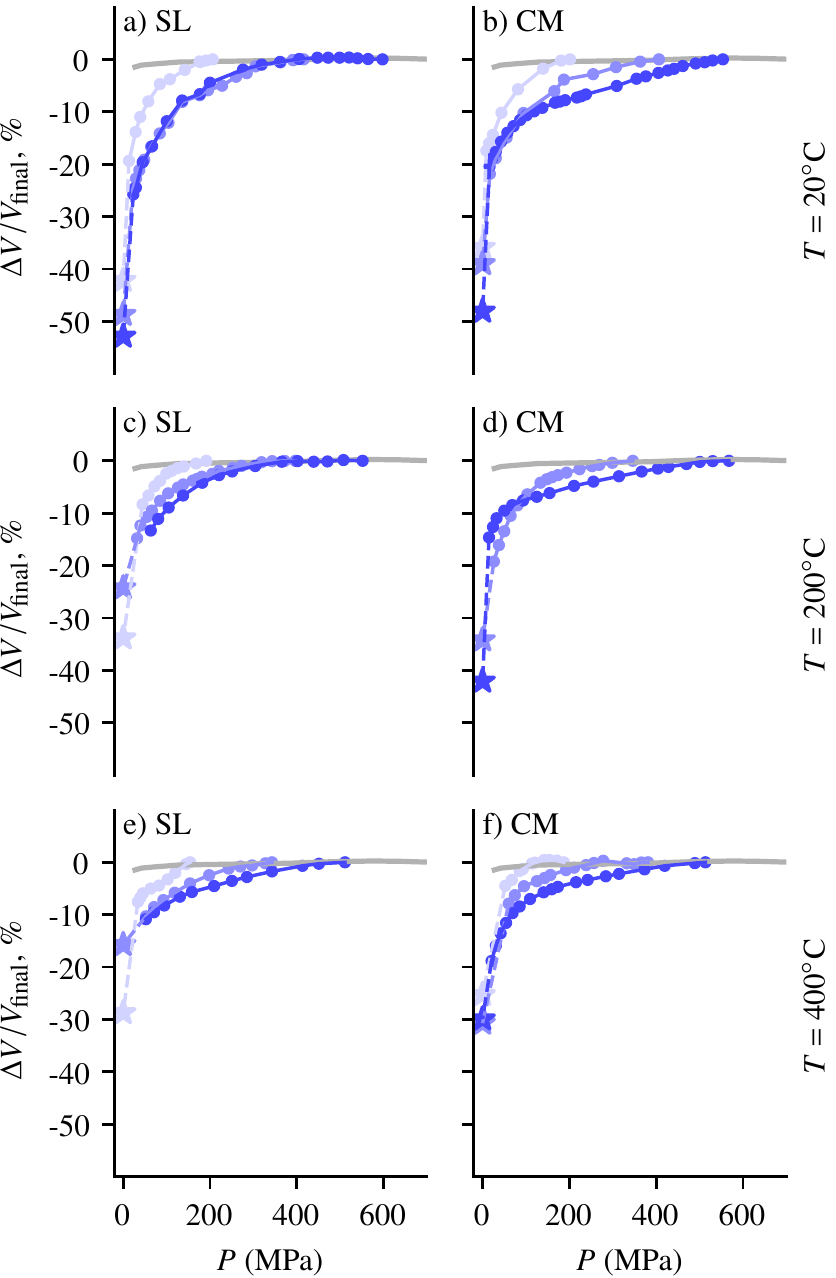}
\caption{Relative change in P-wave speed referenced to the final wave speed ($V_\mathrm{final}$) measured during the unloading phase of each experiment and after sample cooling. Stars denote the wave speed measured at atmospheric pressure after sample recovery. The grey curve represents measurements of wave speed in a fused silica blank over the same pressure range.}
\label{Fig:Decomp}
\end{figure}

Wave speed always decreases significantly during decompression (Figure \ref{Fig:Decomp}). In Solnhofen limestone, the P-wave speed remains approximately constant during initial decompression down to around $400$~MPa. Further decompression leads to a decrease in wave speed, which drops substantially at pressures below $200$~MPa. In samples deformed at room temperature, the wave speed of the recovered material is as low as 50\% of the wave speed measured after deformation but prior to decompression. This drop becomes less marked in samples deformed at high temperature, with changes on the order of 25–35\% at $T=200^\circ$C, and 15–30\% at $T=400^\circ$C. The behaviour is generally similar in Carrara marble, with some notable quantitative differences. The wave speed starts decreasing immediately as pressure in decreased, even in tests conducted at $P=600$~MPa (Figure \ref{Fig:Decomp}b,d). The characteristic pressure below which P-wave speed drops most markedly is on the order of $100$~MPa. The effect of deformation temperature is similar to that observed in Solnhofen limestone, with elevated temperature during deformation promoting a more limited reduction in wave speed during decompression. 

\subsection{Microstructures}

\begin{figure}
    \centering
    \includegraphics{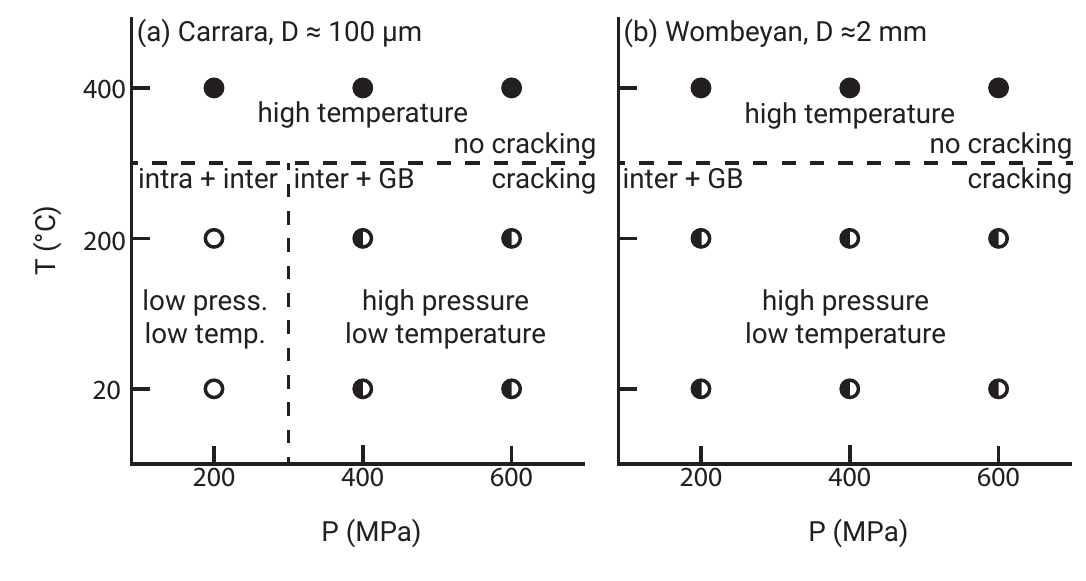}
    \caption{Span of microstructural regimes identified in deformed samples.}
    \label{fig:micro_regimes}
\end{figure}

\subsubsection{Brittle features}

\begin{table}
    \centering
    \caption{Crack densities, $S_v$, in selected samples as the volumetric number of cracks determined from stereological measurements following \citeA{Underwood1970}}.
    \label{table:crack density}
    \begin{tabular}{lcccc}
    \hline
         Sample     &   Lithology   &   Pressure    &   Temperature &   $S_v$ (mm$^2$/mm$^3$)   \\ \hline
         Run0075    &   Carrara     &   200         &   19.8        &   18.73                   \\
         Run0078    &   Carrara     &   405         &   19.8        &   7.31                    \\
         Run0084    &   Carrara     &   615         &   19.8        &   2.28                    \\
         Run0086    &   Carrara     &   236         &   197         &   7.60                    \\
         Run0090    &   Carrara     &   213         &   390         &   2.05                    \\
         Run0093    &   Carrara     &   599         &   192         &   7.82                    \\
         Run0094    &   Carrara     &   594         &   401         &   1.30                    \\
         Run0098    &   Carrara     &   769         &   20.1        &   4.60                    \\
         Run0104    &   Wombeyan    &   206         &   19.3        &   4.02                    \\
         Run0106    &   Wombeyan    &   595         &   20.2        &   0.33                    \\
         Run0110    &   Wombeyan    &   231         &   207         &   1.51                    \\
         Run0119    &   Wombeyan    &   609         &   187         &   0.49                    \\
         Run0120    &   Wombeyan    &   408         &   19.2        &   1.60                    \\
         Run0121    &   Wombeyan    &   428         &   190         &   0.35                    \\
         Run0124    &   Wombeyan    &   233         &   408         &   0.47                    \\
         Run0126    &   Wombeyan    &   574         &   416         &   0.36                    \\\hline
    \end{tabular}
\end{table}

\begin{figure*}[htp!]
    \centering
    \includegraphics{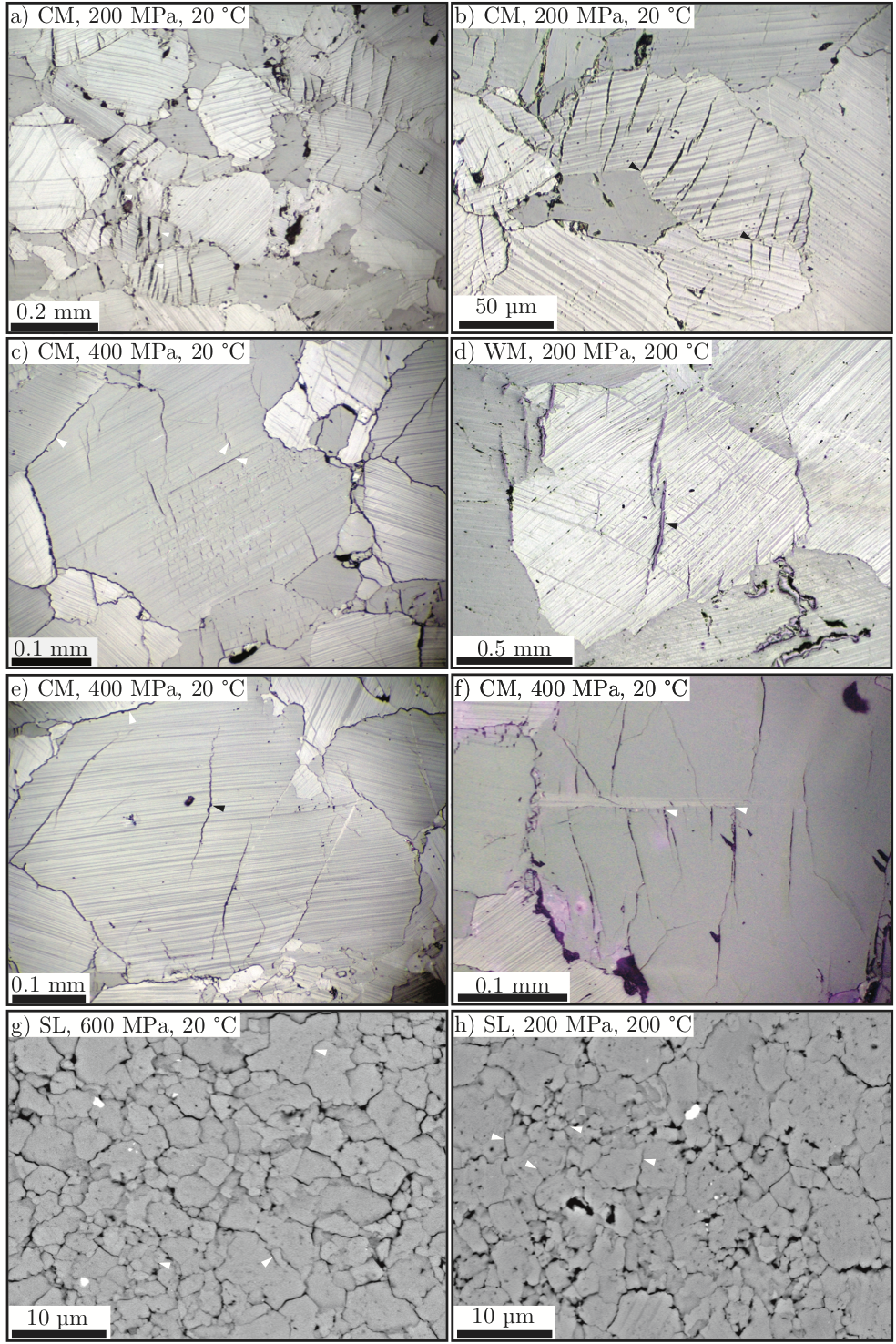}
    \caption{Brittle microstructures observed in samples after testing. a) Crush zones, b) geometrically controlled cracks, c) inter-twin cracks and stair-step cracks, d) grain-boundary cracks and low aspect-ratio cracks, e) stair-step cracks with orientations controlled by twin boundaries and f) cracks nucleated from twins.}
    \label{fig: brittle_micro}
\end{figure*}

Systematic observations of cracks within deformed samples reveals three distinct deformation regimes according to deformation conditions (figure \ref{fig:micro_regimes}). The low-pressure, low-temperature regime and the high-pressure, low-temperature regime are characterised by cracks, whereas in the high-temperature regime cracks are absent.

In the low temperature regimes where cracks are observed, cracks interact with twins. Dense arrays of cracks confined to individual twin lamellae are commonly observed (Figure \ref{fig: brittle_micro}c). In addition, some intragranular cracks contain steps and change orientation across individual twin lamellae ('stair-step', Figure \ref{fig: brittle_micro}e). In other instances, microcracks are observed to nucleate from the tips of twins, and do not reach grain boundaries (Figure \ref{fig: brittle_micro}f).

\noindent
\textbf{Low-pressure, low-temperature regime}

The low-pressure, low-temperature regime is limited to Carrara marble experiments conducted at 200~MPa and 20 and 200$^\circ$C only.  In this regime mode-I intragranular microcracks are a common feature, and result in the highest crack densities. These microcracks are typically confined to single grains and are of low aspect ratio (\textit{e.g.}, Figure \ref{fig: brittle_micro}b,d). In places, the microcracks are concentrated in discrete regions as crush zones (Figure \ref{fig: brittle_micro}a). In these regions, cracks often span a few grains and result in a locally elevated crack density. 

Cracks are also observed to relate to the geometry of grains and grain boundaries. Some cracks nucleate at geometric irregularities along grain boundaries. For example, steps in grain boundaries are often associated with short tensile cracks that propagate a short distance into the grain interior (Figure \ref{fig: brittle_micro}d). Smaller grains can also act as indenter grains, and cracks are nucleated to accommodate the indenter grain shape (Figure \ref{fig: brittle_micro}b). Often, cracks relating to geometric incompatibilities can be wholly contained within a single grain (Figure \ref{fig: brittle_micro}b and d).

\noindent
\textbf{High-pressure, low-temperature regime}

The high-pressure, low-temperature regime spans samples deformed above $P = 200$~MPa and at $T = 20^\circ$ and 200~$^\circ$C for Carrara marble, and all samples deformed at 20$^\circ$C and 200$^\circ$C for Wombeyan marble (Figure \ref{fig:micro_regimes}). This regime is characterised by a reduced crack density ($S_v <$8 mm$^2$/mm$^3$) with respect to the low-pressure, low-temperature regime. Intragranular cracks are not completely suppressed (Figure \ref{fig: brittle_micro}d), e) and f)). However, thorough observations revealed no intergranular cracking and crush zones. Another important change compared to the low-pressure, low-temperature regime is the abundant observation of open grain boundaries (Figure \ref{fig: brittle_micro}c and e).

Within the high-pressure, low-temperature regime, the crack density decreases with increasing pressure for Carrara marble and Wombeyan marble. For example in Carrara marble at 20$^\circ$C, $S_v =$ 7.31~mm$^2$/mm$^3$ at $P =$~400~MPa reducing to $S_v =$ 2.28~mm$^2$/mm$^3$ at $P =$~600~MPa (Table \ref{table:crack density}, Run0078 and Run0084). Furthermore, for a given set of conditions in the high-pressure, low-temperature regime, crack density is lower in the coarser grained Wombeyan marble.

\subsubsection{Crystal-plastic microstructures}

\begin{figure}
    \centering
    \includegraphics{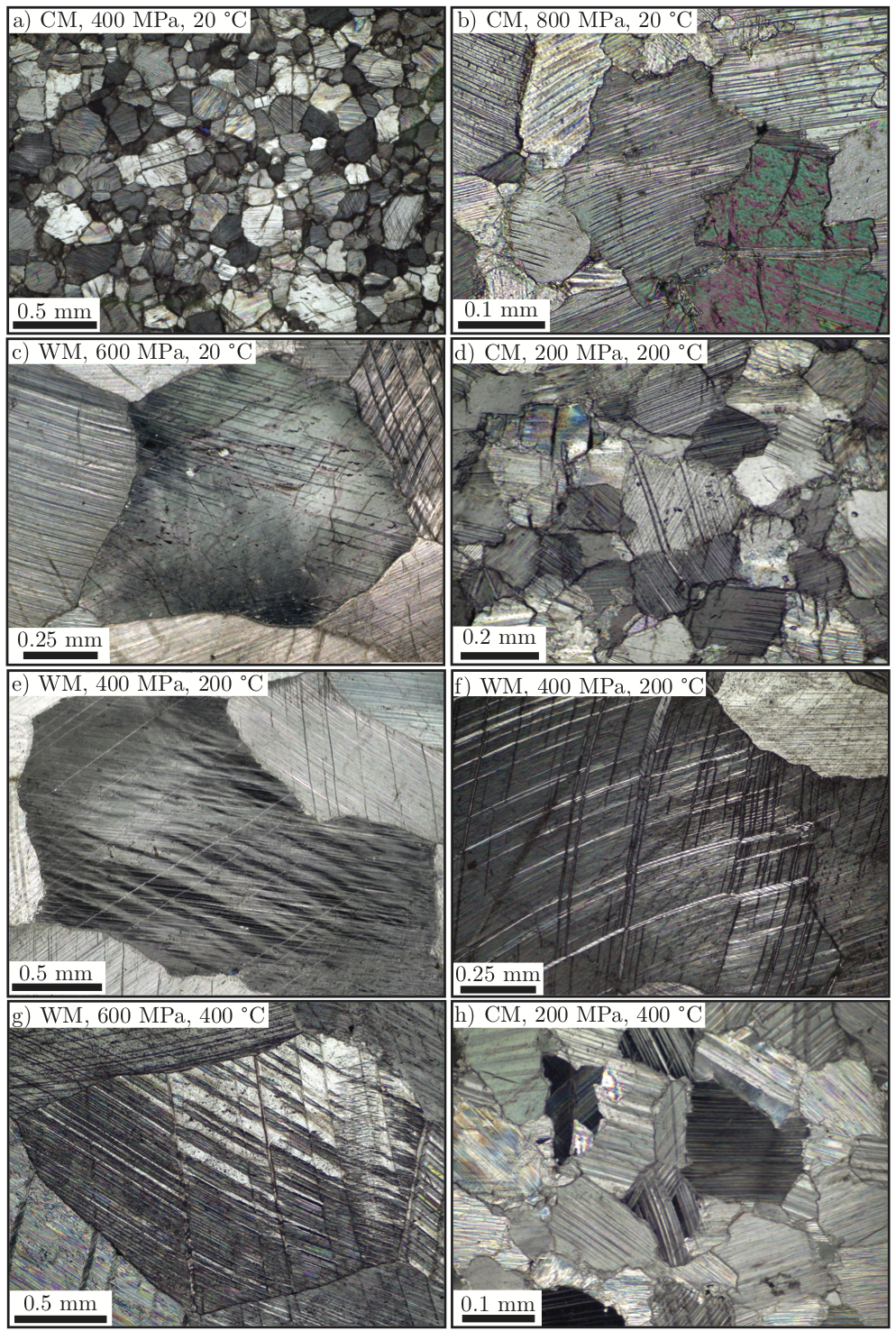}
    \caption{a) High twin density generated at high-stress, low-temperature conditions. b) High-density twins bending around an apparent deformation band. Twins are also nucleated by twins in neighbouring grains. c) Long-range undulose extinction traversing twin sets. The undulose extinction appears to be generated by the interaction of neighbouring grains. d) Multiply-twinned grains and high twin density. e) Corrugation of crystal lattice identified by wavy undulation. f) Multiple curved twin sets and twinned twin sets. g) Multiply-twinned grain in Wombeyan marble. Twins are thicker than at lower-temperature conditions and also appear more patchy. h) Twins formed  at high temperature in Carrara marble. Twins are thicker and the central grain shows evidence of undulose extinction.}
    \label{fig: plastic_micro}
\end{figure}

Crystal-plastic microstructural features are dominated by deformation twins that are present in samples deformed at all conditions (Figure \ref{fig: plastic_micro}). Nearly all grains are twinned on at least one plane, grains that contain two twin sets are also common, and occasionally grains contain three twin sets. In particular, the density of twins, \textit{i.e.}, the number of lamellae per unit length, is high in samples deformed at low temperature (Figure \ref{fig: plastic_micro} a-d) and decreases with increasing temperature (Figure \ref{fig: plastic_micro}g,h). Twins are often curved, especially in the vicinity of grain boundaries (Figure \ref{fig: plastic_micro}f) or around geometric irregularities (Figure \ref{fig: plastic_micro} b). Twins also often appear to have nucleated from twins within neighbouring grains (figure \ref{fig: plastic_micro}b,d) or from geometric irregularities at grain boundaries (figure \ref{fig: plastic_micro}f).

On a larger scale than twin lamellae, undulose extinction is widespread (Figure \ref{fig: plastic_micro}). In Wombeyan marble, the intensity of the undulose extinction increases when approaching grain boundaries and grain-boundary irregularities (Figure \ref{fig: plastic_micro}c). In Wombeyan marble, the undulose extinction observed in some instances suggests corrugation of the crystal lattice (Figure \ref{fig: plastic_micro}e). In Carrara marble, grains also display undulose extinction, although it is not as common as in Wombeyan marble. Again, in Carrara marble undulose extinction is associated to geometric irregularities (Figure \ref{fig: plastic_micro}h).

\subsubsection{SEM observations}

\begin{figure}
    \centering
    \includegraphics{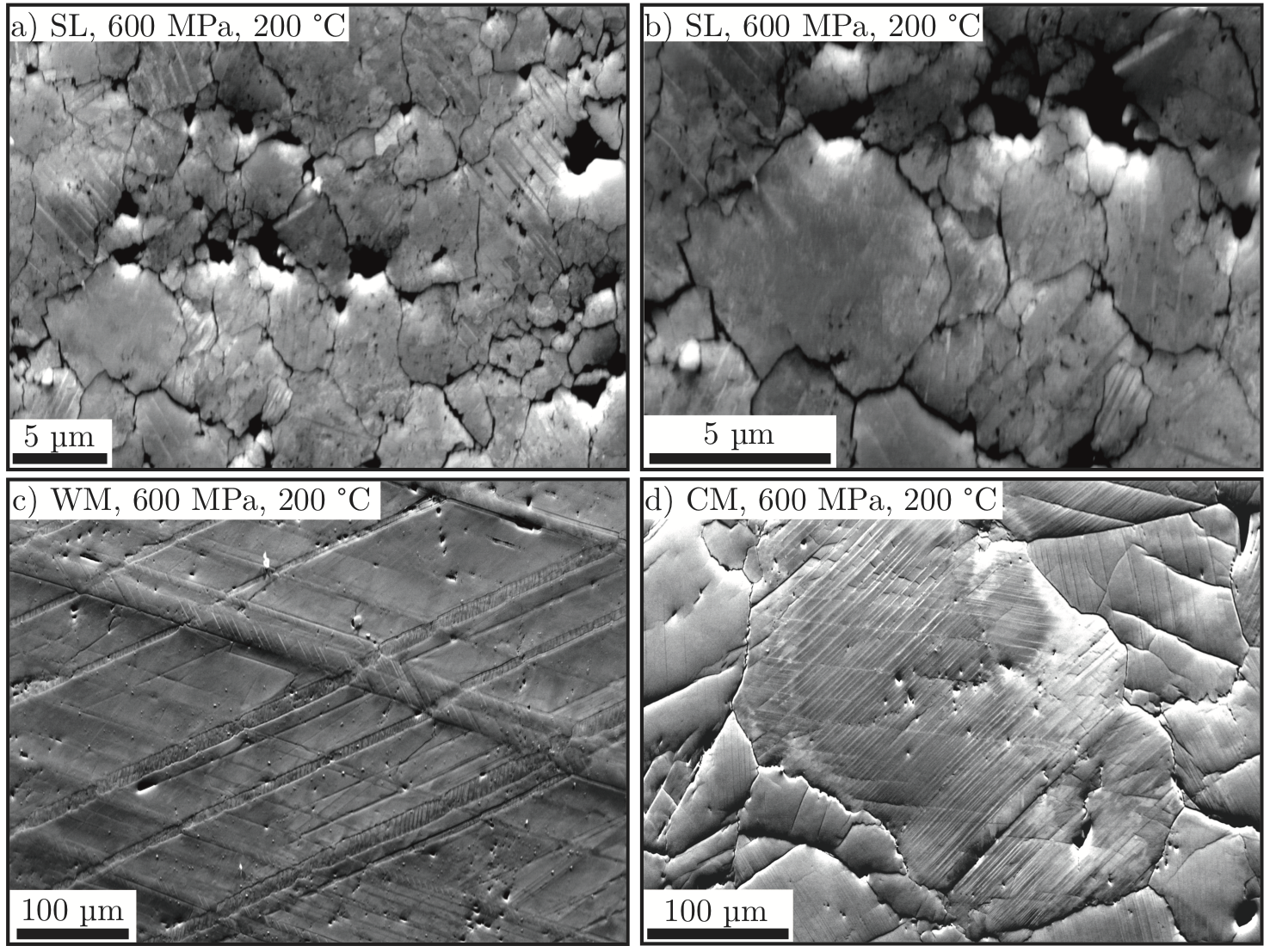}
    \caption{Forescattered-electron orientation-contrast images \cite{Prior1996} of deformation twins in (a and b) Solnhofen limestone, (c) Wombeyan marble and (d) Carrara marble.}
    \label{fig:FS_twins}
\end{figure}

Forescattered electron images were used to image twins at high spatial resolution. These images reveal very thin deformation twins, especially in Solnhofen limestone, which in some places are on the order of 100 nm in thickness (Figure \ref{fig:FS_twins}a,b). Grains of calcite in Solnhofen limestone also exhibit lower twin incidence than the coarser-grained samples of Wombeyan marble and Carrara marble (Figure \ref{fig:FS_twins}c,d). Twins in Solnhofen limestone also propagate across grain boundaries, and are not bent on the grain scale, although they do sometimes taper in proximity to grain boundaries. High-resolution electron images of Wombeyan marble reveal multiple scales of twins, with micrometre-scale twins contained within thicker twins on the order of tens of micrometres in thickness (Figure \ref{fig:FS_twins}c).

\begin{figure}
    \centering
    \includegraphics{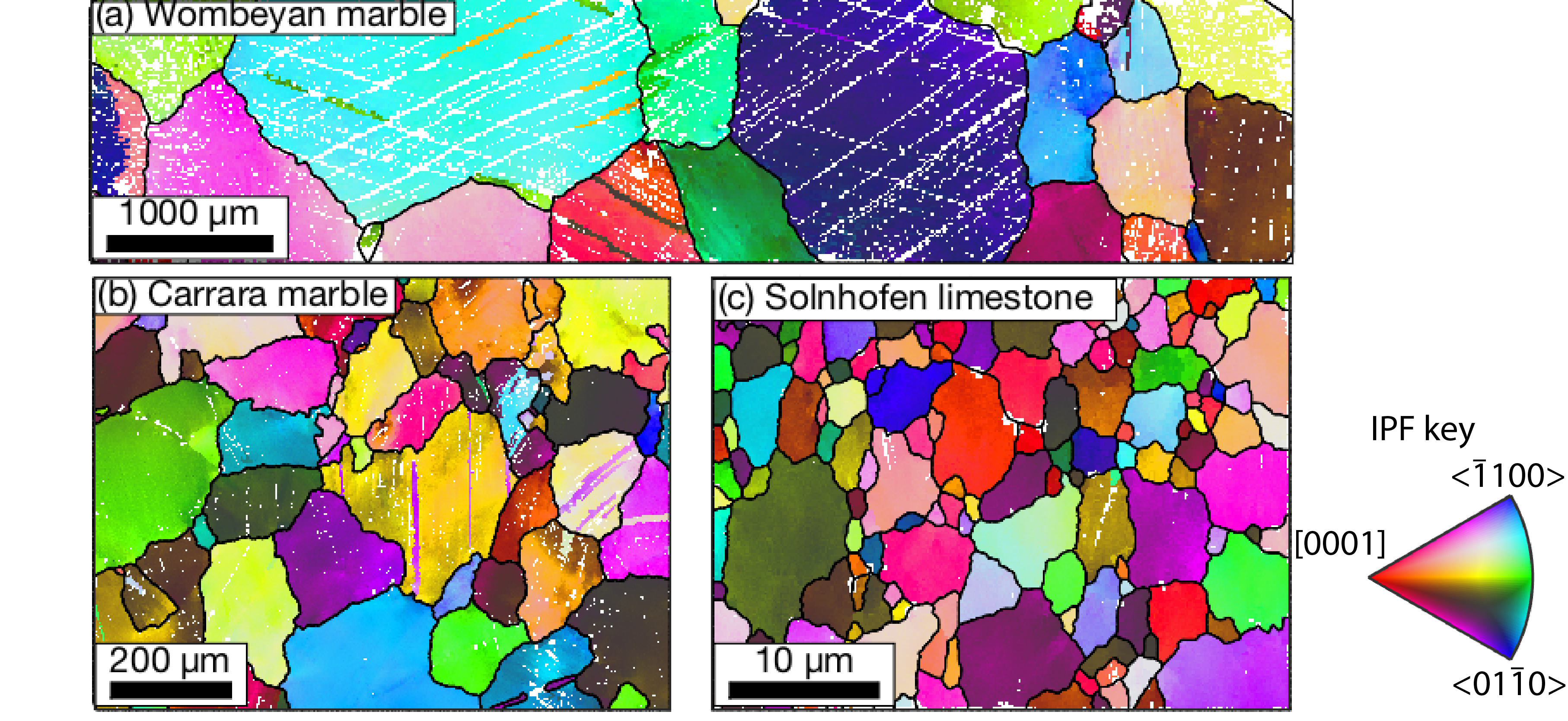}
    \caption{EBSD maps of samples deformed at a pressure of 600 MPa and temperature of 200$^\circ$C. (a) Wombeyan marble, (b) Carrara marble, and (c) Solnhofen limestone. Grains are coloured according to the inverse pole figure. Twin boundaries are not shown for clarity. White areas were not indexed and are mostly due to poor quality diffraction patterns along twin boundaries.}
    \label{fig:EBSD}
\end{figure}

Overview EBSD maps demonstrate that most grains have significant internal misorientation. Lattice curvature is particularly obvious in Wombeyan marble (Figure \ref{fig:EBSD}a) and in Carrara marble (Figure \ref{fig:EBSD}b) and in some grains of Solnhofen limestone (Figure \ref{fig:EBSD}c). Regions of lattice curvature are often related to the geometry of grain boundaries and in places curvature increases in the vicinity of grain boundaries. Other grains exhibit lattice rotation in the vicinity of twin planes, indicated by stripes in the inverse pole figure map (\textit{e.g.}, Figure \ref{fig:EBSD}b). 

\begin{figure}
    \centering
    \includegraphics[width=\textwidth]{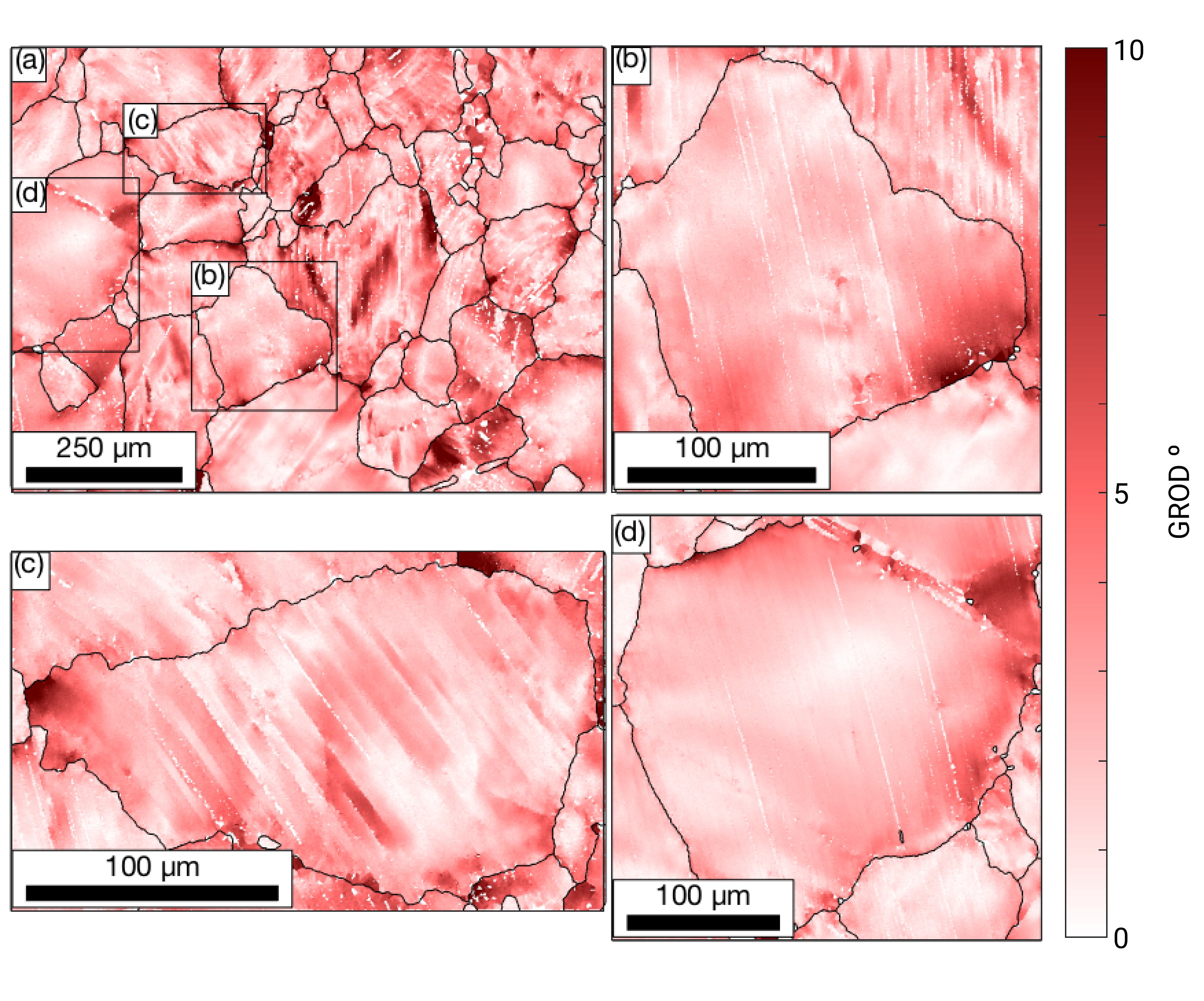}
    \caption{Map of grain reference orientation deviation (GROD) obtained from EBSD maps of Run0093 ($P$ = 600 MPa, $T$ = 200$^\circ$C). Insets (b), (c) and (d) are small-area EBSD maps showing the local GROD within grains with varying interaction between lattice distortion and twins. Strong interactions between twins and lattice curvature are evident in (c), intermediate interaction is present in (b), and weak interaction is apparent in (d). Insets b) and d) also exhibit significant lattice curvature in the vicinity of grain boundaries.}
    \label{fig:GROD}
\end{figure}

Further evidence of internal lattice distortion is revealed by maps of the grain reference orientation deviation (GROD), which is the misorientation of each point with respect to the mean orientation of the grain (Figure \ref{fig:GROD}). An overview map of the GROD in Carrara marble Run0093, deformed at a pressure of $600$~MPa and temperature of $200^\circ$C demonstrates that individual grains have variable internal structure (Figure \ref{fig:GROD}a). Some grains exhibit striped patterns of variable GROD that follow the local twin orientation (Figure \ref{fig:GROD}c). In other grains, the GROD pattern is largely unaffected by twins and instead exhibits lattice distortion over larger length scales approaching the grain size (Figure \ref{fig:GROD}d). In some cases, there is a mixed interaction, where the GROD weakly follows the twin orientation but is also affected by the grain geometry (Figure \ref{fig:GROD}b).

\begin{figure}
    \centering
    \includegraphics[width=\textwidth]{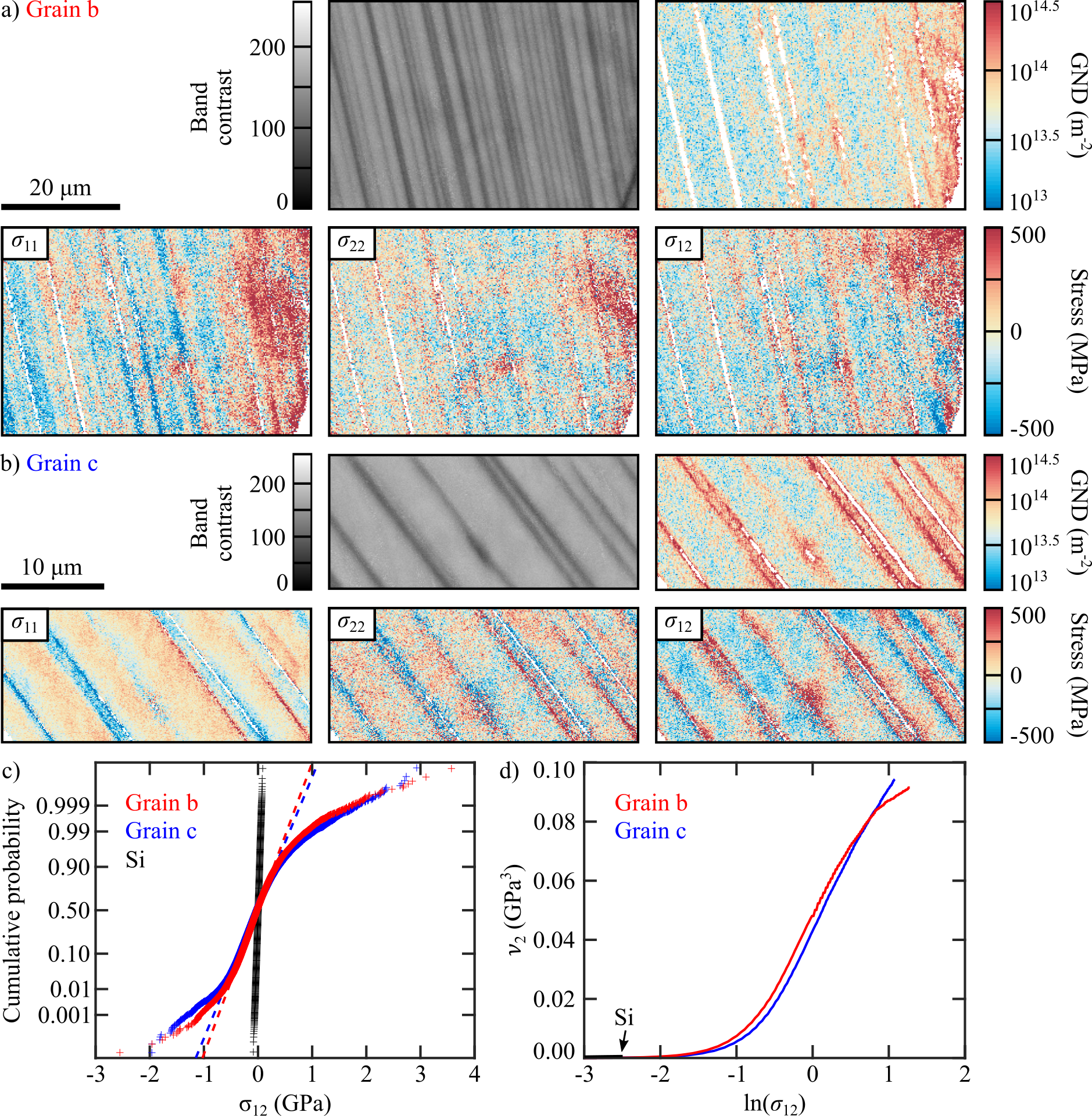}
    \caption{HR-EBSD maps of (a) the grain in Figure \ref{fig:GROD}b and (b) the grain in Figure \ref{fig:GROD}c. Each of these subfigures presents maps of the band contrast in the diffraction patterns, which reveals the locations of twins, GND density, and heterogeneity in stress ($\sigma_{ij}$). (c) Normal probability plot of $\sigma_{12}$ in each grain. Straight lines indicate a normal distribution. (d) Restricted second moment ($\nu_{2}$) versus ln$\sigma_{12}$. Straight lines indicate that the probability distribution of the stress exhibits the form $P(\sigma)\propto|\sigma| ^{-3}$ expected of a population of dislocations.}
    \label{fig:HR_ebsd}
\end{figure}

HR-EBSD maps (Figure \ref{fig:HR_ebsd}) reveal the distributions of GNDs and intragranular stress heterogeneity in the endmember grains in Figure \ref{fig:GROD}. The grain in Figure \ref{fig:GROD}b, which exhibited the least impact of the twins on the distribution of GROD, also exhibits negligible impact of the twins on GND density (Figure \ref{fig:HR_ebsd}a). Within the grain interior, GND density is generally at or below the background noise level of approximately $10^{13.5}$~m$^{-2}$ arising from noise in the rotation measurements. Apparent GND densities above this noise level are highly localised in the vicinity of twin boundaries and potentially result from reduced pattern quality indicated by the band-contrast map. However, GND densities are significantly elevated to the order of $10^{14}$~m$^{-2}$ adjacent to a grain boundary along the right edge of the map. Comparable distributions are evident in the maps of stress heterogeneity with stresses in the grain interior being relatively homogeneous and stresses near the grain boundary being elevated to several hundred megapascals. Different distributions are apparent in HR-EBSD maps (Figure \ref{fig:HR_ebsd}b) from within the grain in Figure \ref{fig:GROD}c, which exhibited the greatest impact of the twins on the distribution of GROD. In this grain, zones of elevated GND density and stress extend a few micrometres from the twin boundaries and beyond the zones of reduced pattern quality represented in the band-contrast map. Within these zones, GND densities approach $10^{14}$~m$^{-2}$ and shear stresses on the order of several hundred megapascals are common.

The probability distributions of the stress heterogeneity provide further information on the cause of the stresses. The normal probability plot (Figure \ref{fig:HR_ebsd}c) exhibits similar probability distributions for both grains, with stresses that are significantly greater than those measured on the undeformed Si standard. Below stress magnitudes of approximately 300–500 MPa, the distributions fall on a straight line indicating these stresses are normally distributed. However, at greater stress magnitudues, the distributions depart from straight lines. These high-stress tails are typical of materials, including Cu and olivine, deformed by dislocation-mediated mechanisms \cite{Jiang2013,Wallis2021,Wallis2022}. The plot of the restricted second moment $\nu_{2}$ versus ln($\sigma_{12}$) provides a further test of whether these high-magnitude stresses are the stress fields of dislocations (Figure \ref{fig:HR_ebsd}d). On this plot, the distributions from the maps from each grain both fall on straight lines at high stresses, indicating that the distributions have the $P(\sigma)\propto|\sigma| ^{-3}$ form that is characteristic of the stress field of a population of dislocations \cite{Groma1998,Wilkinson2014,Kalacska2017,Wallis2021}. These characteristics indicate that the stresses are, at least in part, the stress fields of dislocations.

\begin{figure}
\centering
\includegraphics{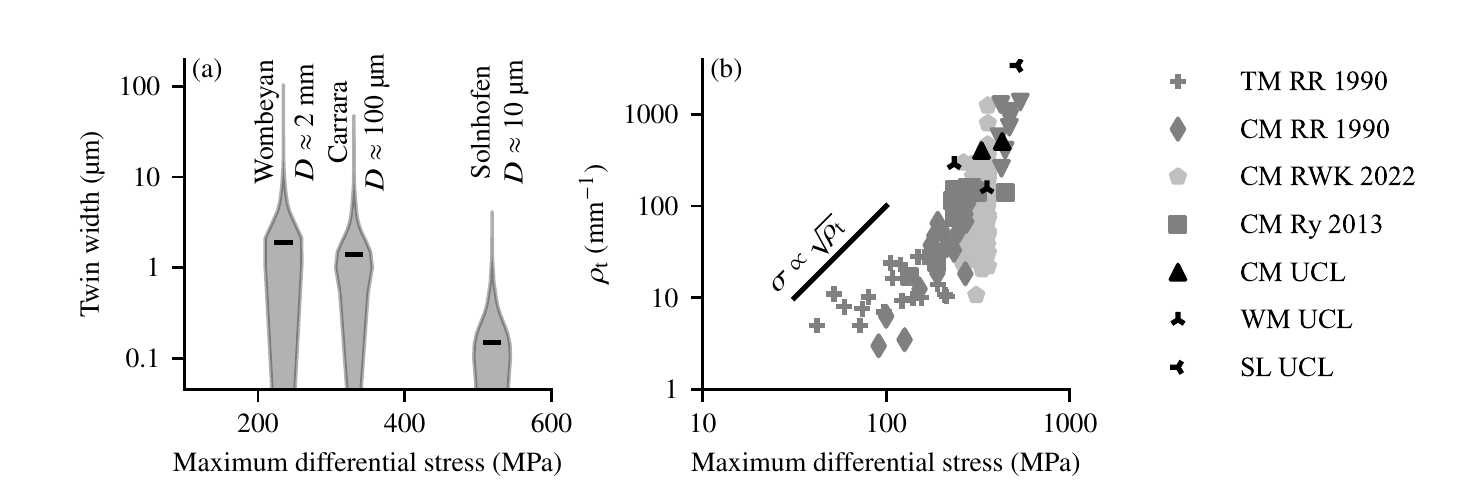}
\caption{The final differential stress at the end of deformation as a function of twin density for this study is shown in black symbols. Additional data from previous studies are shown in grey. CM RR 1990, TM RR 1990 and SL RR 1990 are data from Carrara marble, Taiwan marble, and Solnhofen limestone by \citeA{Rowe1990}, CM RWK 2022 is data from Carrara marble data by \citeA{Rutter2022}, and CM Ry 2013 is data from Carrara marble by \citeA{Rybacki2013}. The reference line indicates the gradient of a relationship in which stress is dependant on the square root of twin density.}
\label{Fig:stress_td}
\end{figure}

In addition to the EBSD mapping, based on the forescatter imaging, we were able to measure twin spacing and correct their thickness in the same manner as \citeA<>[Figure \ref{Fig:stress_td}]{Rutter2022}. We report values of twin density, $\rho_\mathrm{t}$ (mm$^{-1}$), which is computed as the inverse of the measured average twin spacing (Figure \ref{Fig:stress_td}b). Our data agree with previously reported measurements of stress versus twin density. Twin density is lowest in Wombeyan marble deformed at a pressure of 600 MPa and temperature of 200$^\circ$C with a value of 130 mm$^{-1}$, equivalent to a mean spacing of 7.5~$\mu$m. The highest density is obtained for Solnhofen limestone deformed at 600 MPa and 200$^\circ$C, with a value of about 1500 mm$^{-1}$, equivalent to a mean twin spacing of 0.7~$\mu$m.

\section{Discussion}
The mechanical data reveal that deformation of calcite rocks across the range of conditions that we tested is ductile. Almost all experiments demonstrate strain-hardening behaviour but the precise characteristics of the hardening vary with pressure and temperature. The deformation of calcite rocks in this study is accommodated by cracking, twinning and dislocation motion, depending on experimental conditions. The main question we seek to answer is what controls the rheological behaviour of calcite rocks in the semi-brittle regime? Determination of the rheological behaviour of calcite rocks requires knowledge of the approximate partitioning of strain between each active deformation mechanism. This can be answered by considering the microstructural, wave-speed and mechanical data.

\subsection{Brittle deformation}

\begin{figure*}[htp!]
    \centering
    \includegraphics{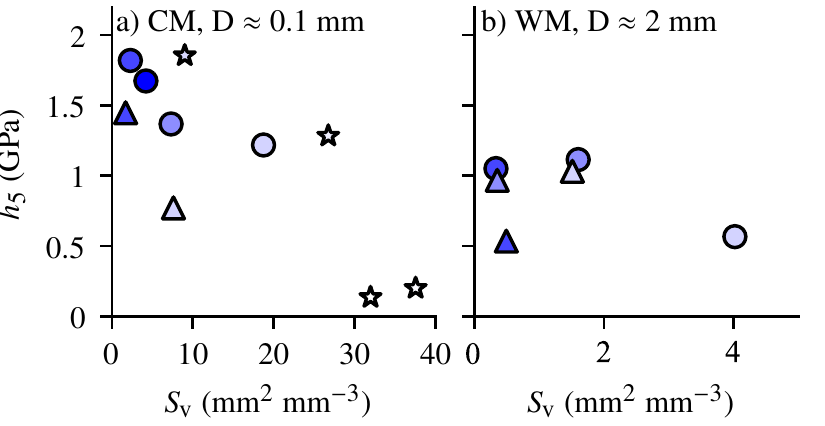}
    \caption{Hardening modulus computed at 5\% strain plotted against post-mortem crack densities. a) Carrara marble data from room-temperature experiments (circles this study, stars \citeA{Fredrich1989}) and at $200^\circ$C (triangles). b) Wombeyan marble from experiments at room temperature (circles) and $200^\circ$C (triangles).}
    \label{Fig:velocity_drop_H}
\end{figure*}

The evolution of wave speed combined with post-mortem measurements of crack density characterise the degree of microcracking occurring during deformation. At all temperatures at which strength is pressure sensitive, increasing pressure results in increasing hardening rates, smaller wave-speed decreases and a reduction in post-mortem crack density (Figure \ref{Fig:velocity_drop_H} and Table \ref{table:crack density}). Similar observations of decreasing \textit{post-mortem} crack density with increasing pressure were made by \citeA{Fredrich1989} in Carrara marble at room temperature. Measurements of volumetric strain during the deformation of Carrara marble and Solnhofen limestone at room temperature made by \citeA{Edmond1972} demonstrate that dilatancy is suppressed at high pressure. Both of these studies also report increased hardening rates with increasing pressure. Decreasing strain accommodation by cracking is therefore systematically associated with increasing hardening rates (Figure \ref{Fig:velocity_drop_H}).

With increasing temperature, pressure-independent strength is reached at a lower pressure. In conjunction with this, deformation at elevated pressure is systematically related to smaller drops in wave speed and increased hardening rates. Crack density is also lower for a given pressure at higher temperature. Measurements of reduced pore volume during deformation of Carrara marble at high temperatures by \citeA{Fischer1989} support this observation. Taken together, these observations identify that cracking is suppressed by temperature increases.

Another interesting observation is the large wave-speed decrease during decompression. This feature suggests that significant sample damage is accrued during decompression. Samples with the largest wave-speed decrease are those deformed at high-pressure, low-temperature conditions, which correspond to a low intragranular fracture density (table \ref{table:crack density}). These samples are characterised by open grain boundaries (Figure \ref{fig: brittle_micro}b, c, e and g), suggesting that grain-boundary opening is the cause of the observed wave-speed decrease. Similar observations were made by \citeA{Edmond1972}, who reported increases in volumetric strain during decompression in deformed Solnhofen limestone and Carrara marble, with the magnitude of volumetric strain change increasing with the final stress level. This phenomenon is likely to result from deformation caused by strain incompatibility between grains \cite{Ashby1970}. During deformation at high pressure, individual grains deform both plastically and elastically to accommodate the imposed strain in the sample, which results in heterogeneous internal stresses. Upon removal of the confining pressure, these internal stresses are no longer in equilibrium with the applied external forces and are likely to generate interface cracks between grains to accommodate the strain incompatibilities accrued during deformation.

\subsection{Grain-size sensitive behaviour}

Our results demonstrate that yield stress and the hardening modulus are functions of grain size.
This result is consistent with the works of \citeA{Olsson1974} and \citeA{Fredrich1990} and other published results at both low temperatures \cite<>[Figure \ref{Fig:D_Strength} and Table \ref{tab:literature}]{Paterson1958,Heard1960,Mogi1964,Donath1971a,Edmond1972,Rutter1974,Fredrich1989,Rybacki2021} and high temperatures \cite{Renner2002}.

Based on experiments at room temperature and moderate pressure, \citeA{Fredrich1990} discussed grain-size strengthening in the model framework of \citeA{Horii1986}. \citeA{Horii1986} solved a wing-crack model coupled to a 'plastic zone', and their results predicted that coarser-grained materials should be more 'brittle', \textit{i.e.}, the ratio of differential stress to confining pressure at the brittle-ductile transition should decrease with increasing grain size. The results of \citeA{Fredrich1990} showed that this ratio was independent of grain size, which they explained by considering that the plastic yield stress scaled as $D^{-1/2}$. This hypothesis is supported by our observations. More generally, yield stress is observed to scale inversely with grain size in metals \cite{Cordero2016} and also in olivine \cite{Hansen2019}.


Grain-size strengthening is a common phenomenon in metals and has received significant attention in the metallurgy literature (see \citeA{Cordero2016} and \citeA{Li2016} for reviews).  Strengthening with decreasing grain size is usually termed the Hall-Petch effect \cite{Hall1951,Petch1953}, and is typically of the form
\begin{linenomath}
 \begin{equation}
     \sigma = \sigma_0 + K D^{-m},
     \label{eq:HP}
 \end{equation}
 \end{linenomath}
where $\sigma_0$ is the intrinsic resistance of a lattice to dislocation motion, $K$ is a material-dependent Hall-Petch coefficient and $m$ is a dimensionless exponent. In the original observations and theoretical arguments of \citeA{Hall1951} and \citeA{Petch1953}, the exponent $m = 0.5$. Values of $m = 0.2–1$ have been reported in metals \cite{Dunstan2014a,Cordero2016,Li2016}.

\begin{table}
\centering
\begin{tabular}{l c c}
\hline
Variable 			&	$\sigma_0$ (MPa)		& $K$ (MPa m$^{-0.5}$)\\ \hline
\multicolumn{3}{l}{$T = 20~^\circ$C}									\\
$\sigma_y$		&	155						& 0.40					\\
$\sigma_{2.5}$	&	244						& 0.73					\\
$\sigma_{5}$	&	274						& 0.87					\\ \hline
\multicolumn{3}{l}{$T = 200~^\circ$C}									\\
$\sigma_y$		&	83						& 0.51					\\
$\sigma_{2.5}$ &	149						& 0.72					\\
$\sigma_5$		&	185						& 0.79					\\ \hline
\multicolumn{3}{l}{$T = 400~^\circ$C}									\\
$\sigma_y$		&	68						& 0.52					\\
$\sigma_{2.5}$ &	91						& 0.86					\\
$\sigma_5$		&	108						& 0.96					\\ \hline
\end{tabular}
\label{tab:HallPetch}
\caption{Results of fitting collated mechanical data with the Hall-Petch relation, with $m=0.5$ (Equation \ref{eq:HP}).}
\end{table}

We fitted equation \ref{eq:HP} to our data using a least-squares regression and setting $m=0.5$ (Table \ref{tab:HallPetch}), similar to \citeA{Renner2002}. Although our results suggest $m$ in the range 0.3--0.6 (Figure \ref{Fig:D_Strength}), fitting the exponent significantly modifies $\sigma_0$ and $K$ and makes comparison among our datasets challenging. Values of the apparent lattice resistance, $\sigma_0$, and the Hall-Petch coefficient, $K$, increase with strain at all temperatures. Values of the lattice resistance are largest at $T=20^\circ$C, with $\sigma_0 = 155–274$~MPa, and are smaller at $T=200~^\circ$C with $\sigma_0 = 83–185$~MPa, further reducing at $400~^\circ$C to $\sigma_0 = 68-108$~MPa. The values for the lattice resistance are consistent with measurements from single-crystal experiments by \citeA{DeBresser1991} and \citeA{Turner1954} at the same temperatures. Values of the Hall-Petch coefficient are largely unaffected by temperature changes. For the reported values of $\sigma_\mathrm{y}$, $K = 0.4–0.52$~MPa~m$^{-0.5}$, at larger strain for values of $\sigma_{2.5}$ the Hall-Petch coefficient increases to $K = 0.72–0.86$~MPa~m$^{-0.5}$ with further increases at large strain of $K=0.79–0.96$~MPa~m$^{-0.5}$ for $\sigma_5$ values. When $K$ is normalised by the product of the shear modulus $G$ and Burgers vector $b$, with $G$~=~35 GPa and $b$~=~0.74 nm, it falls into the range 0.42–1, which is consistent with values typically obtained for BCC and HCP metals \cite{Cordero2016}. In summary, our data combined with literature sources reveal several key features of the Hall-Petch effect in calcite rocks, 1) the Hall-Petch effect is amplified by strain, being weakest at yield and strong after plastic strain, 2) the apparent lattice resistance increases with plastic strain and decreases with temperature, and 3) the Hall-Petch coefficient is independent of temperature.

Extensive reviews by \citeA{Cordero2016} and \citeA{Li2016} summarise proposed physical models of the Hall-Petch effect. \citeA{Li2016} identify four categories of models: (1) the dislocation pile-up model, in which grain boundaries act as obstacles that cause pile-ups until the stress in front of the pile-up is sufficient to  induce yielding of neighbouring grains \cite{Hall1951,Petch1953}; (2) the grain-boundary ledge model, in which grain- and subgrain-boundary irregularities emit forest dislocations that act as obstacles \cite<see>[]{Li1963}; (3) the plastic-strain model, in which the rate of increase in dislocation density with plastic strain is inversely proportional to grain size \cite{Conrad1967a}; (4) the elastic-anisotropy model, in which interactions among elastically anisotropic grains require the introduction of GNDs, and smaller grains have relatively larger strain gradients and greater GND densities \cite{Meyers1982}. All of these models, except that of \citeA{Meyers1982}, arrive at an expression similar to Equation \ref{eq:HP}. The coefficient $K$ always includes the shear modulus, $G$, and the Burgers vector, $b$, as well as other geometric constants, and $m = 0.5$. However, fitting exercises have shown that for a wide range of metals, $m \neq 0.5$ and may be better fit by $m = -1$ or by the relationship $\ln d/d$ \cite{Dunstan2014a,Li2016}.

A grain-size dependence of the yield stress suggests that the plastic-strain model 3 is not appropriate for calcite as a finite amount of macroscopic plastic strain is required to generate the Hall-Petch effect \cite{Conrad1967a,Ashby1970}. As pointed out by \citeA{Hansen2019}, models 1, 2 and 4 also require finite plastic strain, although this may be sufficiently localised for the macroscopic behaviour still appear elastic  prior to macroscopic yielding \cite{Maas2018}. The temperature dependence of the yield stress suggests that the models must include short-range dislocation interactions, as long-range interactions are elastic and therefore largely temperature insensitive \cite{Hansen2019}. 

In model 1, the Hall-Petch effect is attributed to pile-ups of dislocations at grain boundaries, and their role in promoting yield of neighbouring grains \cite{Hall1951,Petch1953}. In this case, larger grains can support longer pile-ups, which generate more intense stress concentrations that promote yield of neighbouring grains. If this model is relevant, we might expect to see pile-ups of dislocations at grain boundaries and localised strain transfer across grain boundaries. \citeA{QuintanillaTerminel2016} use a micro-grid to estimate microscale strains in deformed Carrara marble and document the occurrence of strain transfer across grain boundaries, although this appears to be in broad zones. Further observations using transmission electron microscopy and HR-EBSD are needed to characterise dislocation structures in the vicinity of grain boundaries in calcite to test the relevance of this model. 

Model 2 considers that dislocations are emitted from grain-boundary ledges \cite<see>[]{Li1963}. As materials with finer grain sizes have greater grain-boundary areas, there are more potential sites for generation of dislocations, resulting in a higher dislocation density and therefore stress required for macroscopic plastic deformation. In our samples it is difficult to assess the grain-boundary ledge model. Some evidence may be that undulose extinction is often controlled by the geometry of grain boundaries (Figure \ref{fig: plastic_micro}), indicating that dislocation activity is influenced by grain boundaries, although it does not identify whether grain boundaries act as dislocations sources. 

The elastic-anistropy model of \citeA{Meyers1982} relies on incompatibilities in elastic strain between neighbouring grains. When stress is applied, GNDs form to accommodate the elastic mismatch between grains. Our microstructural observations and wave-speed data may be compatible with this model. Open grains boundaries (Figure \ref{fig: brittle_micro} c, e and g) and large wave-speed decreases during decompression (Figure \ref{Fig:Decomp}) suggest significant relaxation of internal stress during decompression and, in turn, this internal stress may result from elastic mismatch among grains and associated GND formation. EBSD observations also reveal significant lattice curvature (Figure \ref{fig:GROD}) and increases in GND density in the vicinty of grain boundaries (Figure \ref{fig:HR_ebsd}). Furthermore, microscale observations of strain in deformed calcite rocks by \citeA{Spiers1979} and \citeA{QuintanillaTerminel2016} reveal local heterogeneity in finite strain near to grain boundaries. In particular, \citeA{QuintanillaTerminel2016} identify strain heterogeneity on a scale similar to the grain size. \citeA{Wallis2018} observed an increase in misorientation in the vicinity of grain boundaries in naturally deformed calcite rocks, consistent with the presence of plastic strain gradients imparted in response to strain incompatibility among neighbouring grains \cite{Meyers1982}. 

In summary, our observations suggest that either model 1, 2 or 3 maybe applicable to calcite. More systematic microstructural observations are required to discriminate among these models. As a generality, all models predict that the strength is sensitive to the mean free path of dislocations, which is controlled by grain size. Therefore, Hall-Petch models are closely related to Kocks-Mecking-Estrin (KME) single state-variable models of strength, which incorporate the role of dislocation mean free path \cite{Kocks1966,Kocks1976,Mecking1981}. To explore the KME model we must consider the role of twins, which is discussed in the next section.

\subsection{Is TWIP compatible with the deformation of calcite rocks?}

The activity of mechanical twinning in calcite is closely related to the magnitude of differential stress \cite{Jamison1976,Spiers1979,Rowe1990}.
In experiments, \citeA{Rowe1990} demonstrated that the spacing of twins decreases with increasing stress, independent of temperature and grain size. Twins are observed in all our deformed samples, and twin density increases with differential stress consistently with the observations of \citeA{Rowe1990,Rybacki2013,Rutter2022}. \citeA{Rybacki2013} also demonstrated that stress was proportional to the square root of twin density at low stresses ($< 250$ MPa). However, at high stress, the value of stress saturates with respect to twin density and the square root dependence breaks down \cite<>[and Figure \ref{Fig:stress_td}]{Rowe1990}. Transmission Electron Microscope observations of calcite deformation twins indicate that twins often interact with dislocations \cite{Barber1979,Fredrich1989,Rybacki2013}. Given these observations, \citeA{Rybacki2021} argued that twin spacing linearly decreases as dislocation density increases, as flow stress is proportional to the square root of dislocation density \cite{Taylor1934}. These observations suggest that twin spacing either is a consequence of, or controls, the strength of calcite rocks.

\citeA{Rybacki2021} argued that twin spacing directly controls the strain hardening rate, and by extension the strength, by drawing analogy to twinning induced plasticity steels \cite<TWIP,>[]{DeCooman2018}. Models of TWIP originate from high manganese steels, which exhibit high hardening rates (approx. 3\% of the shear modulus) with respect to other steels (approx. 0.05 \% of the shear modulus), as a consequence of mechanical twinning. The mechanism of TWIP originates from the abrupt changes in crystallographic orientation at twin boundaries, which can act as barriers to dislocation motion. Progressive twinning leads to dynamic refinement of the microstructure \cite{DeCooman2018}, in which finer twin spacing reduces the mean free path of dislocations ($\lambda$) and causes a dynamic Hall-Petch effect.

In phenomological models of TWIP in the metallurgical literature, strain hardening is attributed to a dynamic Hall-Petch effect resulting from progressive twinning \cite{Bouaziz2008}.
The model of \citeA{Bouaziz2008} is formulated by first considering the Taylor equation that relates shear flow stress ($\tau$) to the total dislocation density ($\rho$),
\begin{linenomath}
\begin{equation}
    \tau = \tau_0 + \tau_b + \alpha \mu b \sqrt{\rho},
    \label{eq:taylor}
\end{equation}
\end{linenomath}
where $\tau_0$ is the initial strength of a polycrystal, $\tau_b$ is the back stress, which may arise from long-range dislocation interactions and stress fields around twins, $\alpha$ a constant close to unity, $b$ the Burgers vector and $\mu$ the shear modulus. In their formulation, the final term represents isotropic hardening due to short-range dislocation interactions. In calcite, the Taylor equation has been demonstrated to apply to calcite rocks deformed at temperatures of 550–700$^\circ$C \cite{DeBresser1996}, although the relative contributions of kinematic hardening due to long-range dislocation interactions that generate back stress and isotropic hardening due to short-range dislocation interactions have not been separated.\\

To obtain the evolution of stress with strain, the Taylor relation is combined with a modified Kocks-Mecking-Estrin equation \cite{Kocks1966,Kocks1976,Mecking1981} to describe the change of $\rho$ with strain \cite{Bouaziz2008}:
\begin{linenomath}
\begin{equation}
    \frac{d \rho}{d \varepsilon} = \frac{1}{b \lambda} - f \rho = \frac{1}{b}\left(\frac{1}{D} + \frac{1}{D_\mathrm{t}} + k\sqrt{\rho}\right) - f \rho,
    \label{eq:TWIP}
\end{equation}
\end{linenomath}
where $f$ is a rate- and temperature-dependent dynamic-recovery coefficient, $k$ is a constant that characterises dislocation storage due to dislocation interactions, and $D_t$ is the twin spacing. In this model, changes in the dislocation mean free path are sensitive to the total dislocation density, grain size and twin spacing. An additional term, given by the product of the recovery factor and the dislocation density, $f \rho$, is subtracted to account for dynamic recovery processes.
\citeA{Rybacki2021} argued that this model could potentially capture the rheological behaviour of calcite polycrystals.

The dynamic Hall-Petch model of \citeA{Bouaziz2008} provides a physical basis for the dependence of stress on the square root of twin density at low stress.
\citeA{Rybacki2021} also argued that the high hardening rates (3–5 \% of the shear modulus, Figure \ref{Fig:Hardening}) observed in calcite rocks are also consistent with TWIP. However, it should be noted that high hardening rates (compared to typical expectations in metals of 0.5-1\% of the shear modulus) are not unique to calcite rocks as olivine, which does not twin, exhibits hardening rates up to 5–10 \% of the shear modulus when deformed by low-temperature plasticity \cite{Hansen2019,Druiventak2011}.

The TWIP model also suggests that the hardening rate should be dominated by twin spacing as $D_\mathrm{t}$ is always at least 1–2 orders of magnitude smaller than the grain size $D$ (Fig. \ref{Fig:stress_td}). Taking the case of Carrara marble at 600 MPa, 200 $^\circ$C with $D =$ 100 \textmu m and $D_\mathrm{t} =$ 5 \textmu m, we might expect strain hardening of  $450 \alpha \mu$ (neglecting recovery and forest hardening). At the same conditions for Wombeyan marble, $D = 1$ mm and $D_\mathrm{t} =$ 7.5 \textmu m, so that strain hardening should be of $360 \alpha \mu$. The ratio of hardening rates would therefore be ~1.25. The actual hardening rates observed in our experiments are in a ratio of 2 (considering $H_5 = $ 1.6 and 0.8 GPa for Carrara marble and Wombeyan marble respectively), which suggests that twins may have a smaller impact on hardening than anticipated from Equation \ref{eq:TWIP}.

\subsubsection{Twins as potential barriers to dislocations}

Our main observation is that of grain size strengthening (Figure \ref{Fig:D_Strength}), and it is difficult to assess the exact role of twins in the hardening behaviour from our microscopic data alone. The TWIP model is founded on the notion that twins produce further hardening by either retarding or stopping dislocation motion. In order to test the potential validity of the TWIP model in calcite, in this Section we assess the respective efficiency of grain boundaries and twins at impeding dislocations.

Microstructural observations suggest that the interaction between dislocations and twins boundaries varies between grains. Some grains contain lattice curvature that is clearly affected by twin boundaries (\textit{e.g.}, Figure \ref{fig:GROD}c) whereas in other grains the lattice curvature appears to be affected by twin boundaries (e.g., Figure \ref{fig: plastic_micro}c and Figure \ref{fig:GROD}d). The trapping of dislocations therefore appears to depend on grain orientation and related twin orientation.

We can assess the effectiveness of twin boundaries and grain boundaries as barriers to dislocations by considering slip-transmission coefficients. The simplest form of this analysis is purely geometric and considers only the slip-plane orientation and direction of the Burgers vector of the incoming and outgoing slip systems \cite{Luster1995}, 
\begin{linenomath}
\begin{equation} \label{eq:luster}
    m^\prime = (n_\mathrm{A} \cdot n_\mathrm{B}) (b_\mathrm{A} \cdot b_\mathrm{B}) = \cos(\phi) \cos(\kappa)
\end{equation}
\end{linenomath}
where $n$ denotes the unit normal vector of the slip plane and $b$ the unit Burgers vector, and the subscripts $\mathrm{A}$ and $\mathrm{B}$ denote the incoming and outgoing slip systems. A value of zero for $m^\prime$ indicates an impenetrable barrier to dislocations and a value of one indicates a transparent boundary.

\begin{figure*}[htp!]
    \centering
    \includegraphics{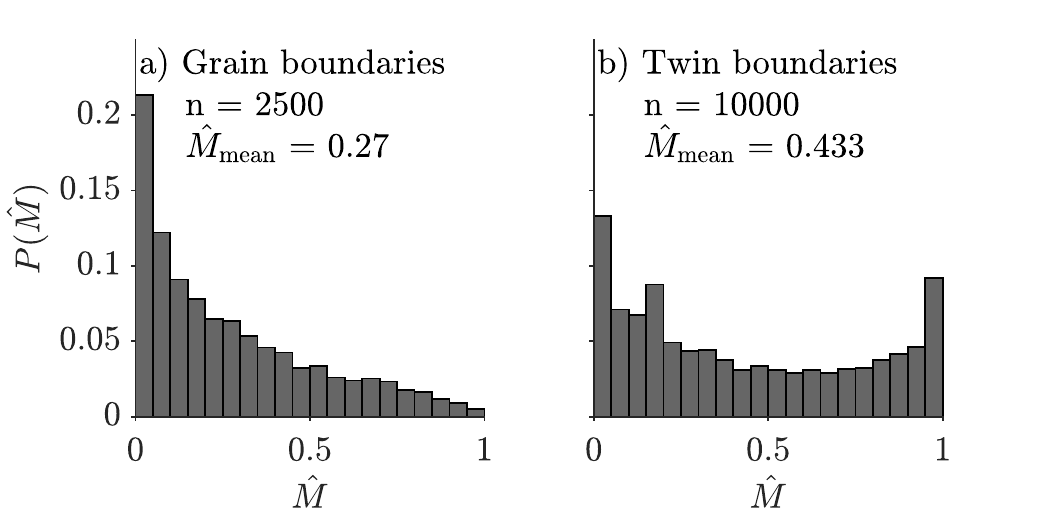}
    \caption{Computed values of $M^\prime$ for a) grain boundaries and b) twin boundaries. }
    \label{fig:slip transmission}
\end{figure*}

Extensive work by \citeA{DeBresser1990,DeBresser1993,DeBresser1997} demonstrated that $r$\hkl{10-14}\hkl<-2021>$^\pm$ and $f$\hkl{-1012}\hkl<2-201>$^\pm$ are the dominant slip systems in calcite at low temperatures. The critical resolved shear stresses of these slip systems is about an order of magnitude greater than that of $e$ twinning at room temperature. Despite their strength, we consider them to be active in our samples since our stress level is typically significantly above this level and we observe significant intragranular misorientation. We computed slip transmission across twin and grain boundaries using a random fabric in MTEX. Maximum values of $m^\prime$ across twin boundaries (Table \ref{tab:literature}) demonstrate that $m^\prime$ is always greater than 0.4, and in some cases twins are transparent with a value of 1 (\textit{e.g.}, $r_2^+$ to $r_3^+$ across an $e_1$ twin). Given also the large number of available slip systems, this analysis suggests that twin boundaries may impede dislocation motion less than randomly oriented grain boundaries.

\begin{table*}[htp!]
    \centering
    \begin{tabular}{c c c c c c c}
    \hline
        $S_\mathrm{A}$          & maximum $m_{e1}^\prime$       & $S_\mathrm{B}$            & maximum $m_{e2}^\prime$   &   $S_\mathrm{B}$          & maximum $m_{e3}^\prime$   & $S_\mathrm{B}$            \\ \hline
        $r_1^-$                 & 0.617                         & $r_1^-$                   & 0.414                     &   $f_1$\hkl<-220-1>$^-$   & 0.414                     & $f_2$\hkl<20-2-1>$^-$     \\
        $r_2^-$                 & 0.414                         & $f_2$\hkl<20-2-1>$^-$     & 0.617                     &   $r_2^-$                 & 0.414                     & $f_2$\hkl<0-22-1>$^-$     \\
        $r_3^-$                 & 0.414                         & $f_3$\hkl<20-2-1>$^-$     & 0.414                     &   $f_3$\hkl<-220-1>$^-$   & 0.617                     & $r_3^-$                   \\
        $r_1^+$                 & 0.630                         & $r_1^+$                   & 1.000                     &   $r_3^+$                 & 1.000                     & $r_2^+$                   \\
        $r_2^+$                 & 1.000                         & $r_3^+$                   & 0.630                     &   $r_2^+$                 & 1.000                     & $r_1^+$                   \\
        $r_3^+$                 & 1.000                         & $r_2^+$                   & 1.000                     &   $r_1^+$                 & 0.630                     & $r_3^+$                   \\
        $f_1$\hkl<-220-1>$^-$   & 0.541                         & $r_1^-$                   & 0.718                     &   $f_1$\hkl<-220-1>$^-$   & 0.596                     & $f_1$\hkl<0-22-1>$^-$     \\
        $f_1$\hkl<2-201>$^+$    & 0.541                         & $r_1^-$                   & 0.718                     &   $f_1$\hkl<-220-1>$^-$   & 0.596                     & $f_1$\hkl<0-22-1>$^-$     \\
        $f_2$\hkl<0-22-1>$^-$   & 0.596                         & $f_2$\hkl<20-2-1>$^-$     & 0.541                     &   $r_2^-$                 & 0.718                     & $f_2$\hkl<0-22-1>$^-$     \\
        $f_2$\hkl<02-21>$^+$    & 0.596                         & $f_2$\hkl<20-2-1>$^-$     & 0.541                     &   $r_2^-$                 & 0.718                     & $f_2$\hkl<0-22-1>$^-$     \\
        $f_3$\hkl<20-2-1>$^-$   & 0.718                         & $f_3$\hkl<20-2-1>$^-$     & 0.596                     &   $f_3$\hkl<-220-1>$^-$   & 0.541                     & $r_3^-$                   \\
        $f_3$\hkl<-2021>$^+$    & 0.718                         & $f_3$\hkl<20-2-1>$^-$     & 0.596                     &   $f_3$\hkl<-220-1>$^-$   & 0.541                     & $r_3^-$                   \\
        $f_1$\hkl<0-22-1>$^-$   & 0.541                         & $r_1^-$                   & 0.596                     &   $f_1$\hkl<-220-1>$^-$   & 0.717                     & $f_1$\hkl<0-22-1>$^-$     \\
        $f_1$\hkl<02-21>$^+$    & 0.541                         & $r_1^-$                   & 0.596                     &   $f_1$\hkl<-220-1>$^-$   & 0.717                     & $f_1$\hkl<0-22-1>$^-$     \\
        $f_2$\hkl<20-2-1>$^-$   & 0.596                         & $f_2$\hkl<20-2-1>$^-$     & 0.541                     &   $r_2^-$                 & 0.596                     & $f_2$\hkl<0-22-1>$^-$     \\
        $f_2$\hkl<-2021>$^+$    & 0.596                         & $f_2$\hkl<20-2-1>$^-$     & 0.541                     &   $r_2^-$                 & 0.596                     & $f_2$\hkl<0-22-1>$^-$     \\
        $f_3$\hkl<-220-1>$^-$   & 0.718                         & $f_3$\hkl<20-2-1>$^-$     & 0.718                     &   $f_3$\hkl<-220-1>$^-$   & 0.541                     & $r_3^-$                   \\
        $f_3$\hkl<2-201>$^+$    & 0.718                         & $f_3$\hkl<20-2-1>$^-$     & 0.718                     &   $f_3$\hkl<-220-1>$^-$   & 0.541                     & $r_3^-$                   \\ \hline
    \end{tabular}
    \caption{Slip-transmission analysis of twin boundaries. Tabulated results for the maximum value of $m^\prime$ (Equation \ref{eq:luster}) between slip system $S_\mathrm{A}$ and $S_\mathrm{B}$. The $m^\prime$ subscript denotes the twinning system considered: \textit{e.g.}, $m_{e1}^\prime$ represents slip transfer across an $e1$ twin.}
    \label{tab:mprime}
\end{table*}

To further compare the effect of twin and grain boundaries on dislocation motion, we can also consider the effects of twin-boundary orientation. This analysis can be performed with a geometric criterion that quantifies the degree of misalignment between the line intersections of incoming and outgoing slip planes with the twin plane, $\mathbf{l}_\mathrm{A}$ and $\mathbf{l}_\mathrm{B}$, respectively, and between their slip directions, $\mathbf{d}_\mathrm{A}$ and $\mathbf{d}_\mathrm{B}$, respectively. The following scalar quantity is maximal when the two slip systems on either side of the boundary are aligned and slip can be transmitted easily across the boundary \cite{Shen1986, Bayerschen2016}:
\begin{linenomath}
\begin{equation}
    \hat{M} = (\mathbf{l}_\mathrm{A} \cdot \mathbf{l}_\mathrm{B})(\mathbf{d}_\mathrm{A} \cdot \mathbf{d}_\mathrm{B}).
\end{equation}
\end{linenomath}
The line intersections $\mathbf{l}$ can be obtained from $\mathbf{l} = (\mathbf{n} \times \mathbf{n}_\mathrm{\Gamma})/| \mathbf{n}_A \times \mathbf{n}_\mathrm{\Gamma} |$, where $\mathbf{n}_\mathrm{\Gamma}$ is the twin boundary plane normal.

We used the $\hat{M}$ criterion to compare the efficiency of slip transmission across twin boundaries to slip transmission across grain boundaries. A random fabric was generated using MTEX, and we computed $\hat{M}$ between random grain pairs by assuming that the activated slip system (either $r^\pm$ or $f^\pm$ slip) was that with highest Schmid factor in each grain. For each grain, we also computed $\hat{M}$ across a twin boundary hosted in the initial grain, the activated twin system was assumed to be that with the highest Schmid factor in each grain. These calculations suggest an average value for slip transmission of $\hat{M} = 0.27$ across grain boundaries, which is considerably smaller than the average value of $\hat{M} = 0.54$ for slip transfer across twin boundaries. 

\begin{figure*}[htp!]
    \centering
    \includegraphics{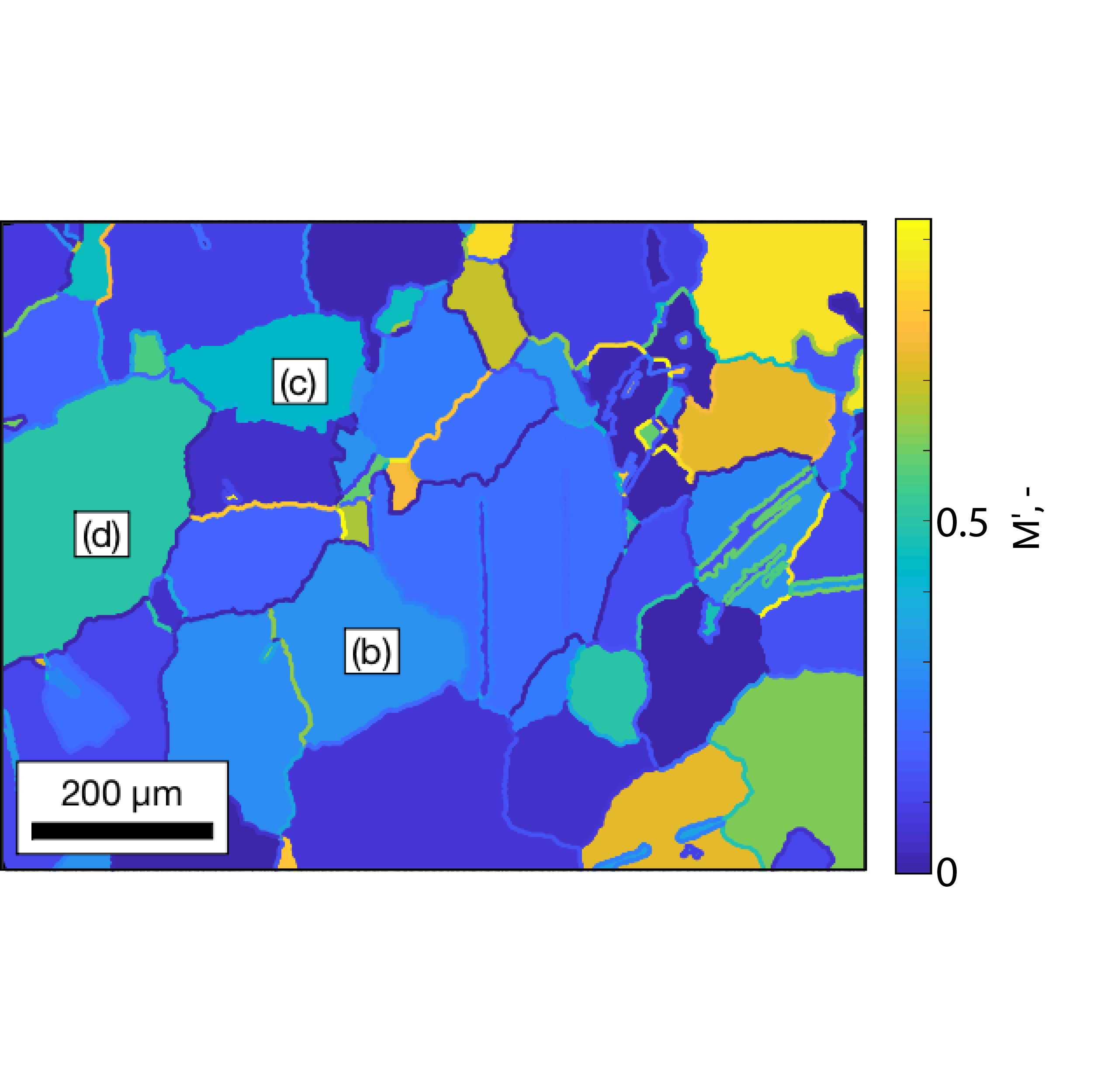}
    \caption{Map of $\hat{M}$ values for slip transmission across grain boundaries (coloured according to value) and twin boundaries (grains coloured according to value) computed from EBSD data obtained for Run0093. } 
    \label{fig:mprime}
\end{figure*}

To further test this result, we computed the expected $m^\prime$ values for our EBSD data across the activated twin systems within each grain (Figure \ref{fig:mprime}). The results demonstrate that grains with low values of $m^\prime$ (Figure \ref{fig:mprime}, grain c) exhibit dislocation substructures that are affected by twinning, that is segmentation of lattice distortion between twins (Figure \ref{fig:GROD}c). In contrast, grains with large values of $m^\prime$ (Figure \ref{fig:mprime}, grain d) exhibit gradients in lattice distortion over length scales approaching the grain size, but which are not strongly affected at smaller scales by twin boundaries. HR-EBSD maps taken from the interior of these grains supports this conclusion, large residual stresses and GND densities are observed in the vicinity of twin boundaries in grains for which $m^\prime$ is small (Figure \ref{fig:HR_ebsd}b and c). Lower residual stresses and GND densities are present in the vicinity of twin boundaries in grains for which $m^\prime$ is low (Figure \ref{fig:HR_ebsd}a and Figure \ref{fig:mprime}b). The efficacy of twin boundaries as barriers to dislocation motion is therefore dependent on the local orientation of individual grains, but is on average weaker than a grain boundary taken at random.

The relative ease with which dislocations can transmit across twin boundaries suggests that Equation \ref{eq:TWIP} requires refinement. We suggest that weights should be applied to the contributions of twin boundaries and grain boundaries, such that equation \ref{eq:TWIP} becomes,
\begin{linenomath}
\begin{equation}
    \frac{d \rho}{d \varepsilon} = \frac{1}{b \lambda} - f \rho = \frac{1}{b}\left(\frac{k_D}{D} + \frac{k_\mathrm{t}}{D_\mathrm{t}} + k\sqrt{\rho}\right) - f \rho,
\end{equation}
\end{linenomath}
in which $k_\mathrm{D}$ and $k_\mathrm{t}$ are weights to account for the relative efficacy of grain boundaries and twin boundaries. We expect that $k_\mathrm{t}\ll k_\mathrm{D}$. Further microstructural measurements, such as the evolution of twin density with strain, are required to determine the value of these weighting factors.

\subsection{Towards a model of semi-brittle flow in calcite rocks}

Our observations combined with previous results suggest several key characteristics that should be captured by a model of semi-brittle flow in calcite-rich rocks at $T<400^\circ$C: (1) non-linearly increasing strength and hardening with increasing pressure, (2) decreasing strength with increasing temperature, (3) increasing strength with decreasing grain size, (4) strength that is nearly insensitive to strain rate. \citeA{Rybacki2021} reviewed proposed models of semi-brittle flow, identifying difficulties in combining brittle and plastic flow into a simple unified model.

Through the semi-brittle flow regime, the strain contribution of brittle and crystal-plastic processes changes with depth. At low pressures and temperatures, at which brittle processes dominate, the macroscopic behaviour of rocks is described by frictional failure \cite{Byerlee1966,Brace1980}. Microphysical models of brittle deformation are typically based the wing-crack model \cite<e.g.,>[]{Nemat-Nasser1982,Ashby1990}, in which brittle damage is accounted for by the propagation of Mode I wing cracks. In this regime, strength is pressure sensitive, dependent on grain size and is to first order strain rate-insensitive. There is no strong temperature dependence, and plasticity is not considered.


There are only a small number of microphysical models in existence accounting for coupled brittle-plastic deformation. \citeA{Horii1986} modified and solved the problem of a wing crack coupled a 'plastic zone' by considering a dislocation pile-up ahead of the shear crack. Their model is sensitive to pressure, grain size and also temperature as the plastic yield strength in the plastic zone can vary with temperature. However, their model predicts that materials with coarser grain size are more brittle, contradicting observations from experimental data \cite{Fredrich1990}.

More recently, \citeA{Nicolas2017} derived a model incorporating the propagation of wing cracks, plastic pore collapse and nucleation of new cracks due to dislocation pile-ups. The model reproduces several important characteristics of the deformation of porous limestones. As pointed out by \citeA{Rybacki2021}, this model does not consider dynamic recovery or twinning, which are important deformation processes in calcite-rich rocks. The large number of free parameters make this model challenging to test and may limit its general application.

An alternative approach to introduce feedbacks between cracking and plastic flow may be by use of a modified Kocks-Mecking equation. The anticorrelation between strain hardening and crack density indicated by our results (Figure \ref{Fig:velocity_drop_H}) suggests that cracking acts to reduce stress and hardening. The role of cracks could be multiple. One possibility is that strength remains dictated by dislocation density, and correctly predicted by the Taylor equation \ref{eq:taylor}, in which case a decreasing flow stress would imply a reduction of dislocation density and thus that and cracks could act as dislocation sinks. This possibility is compatible with the idea that tensile cracks correspond to free surfaces within the material, and dislocations intersecting those free surfaces would form steps and disappear from the crystals. One other possibility is that cracks relax internal stress and strain incompatibilities between grains, i.e., act as "geometrically necessary" structures. A third option is that deformation at low pressure, at which cracks are pervasive, is not controlled by dislocation motion but dominated by elastically-accommodated intergranular slip, and tensile cracks relax the associated internal stresses. 

The origin of microcracks during semi-brittle deformation of calcite is potentially coupled to intracrystalline plasticity. As discussed extensively by \citeA{Nicolas2017}, cracks can be nucleated due to stress concentrations at the head of dislocation pile-ups \cite<e.g.,>[]{Stroh1954,Olsson1976,Wong1990} and may therefore be dislocation sinks, i.e., microcracks could contribute to dislocation escape from the deformed crystals. Where slip transfer is inefficient \cite{Olsson1976} and also where geometry results in a high density of GNDs (e.g., Figure \ref{fig: brittle_micro}d) stresses are high, which can lead to the nucleation of cracks. 

An approach based on the Kocks-Mecking model has the advantage that twinning and grain size effects can be incorporated, whilst introducing a confining-pressure dependence due to the propagation of Mode I cracks. At this stage however, there are insufficient data on dislocation-density evolution during semi-brittle deformation to make sensible progress beyond the qualitative statements listed above. Thus, detailed modelling attempts are beyond the scope of the present work.

\section{Conclusions}
We performed a series of triaxial deformation experiments using three calcite rocks of variable grain size: Solnhofen limestone, Carrara marble and Wombeyan marble. Our experimental results demonstrate that the strength and hardening rates of calcite rocks deformed in the semi-brittle regime are inversely dependent on grain size. 

Microstructural observations using visible-light and electron microscopy demonstrate that strain is accommodated by cracking, twinning and dislocation glide. Deformation tests were accompanied by \textit{in-situ} measurements of P-wave speed, which generally decreases with strain. Wave speed decreases more at room temperature than at $200~^\circ$C and $400~^\circ$C, which is consistent with microcracking being more prevalent at low temperature. 

Quantitative microstructural observations reveal that microcrack density is inversely proportional to hardening rate. While the exact role of cracks in the overall stress-strain behaviour remains unclear at this stage, we propose the hypothesis that tensile microcracks cause weakening by either relaxing internal stresses (accommodating strain incompatilibities and reducing the need for geometrically necessary dislocations), or offering free surfaces where dislocations can escape individual grains, or a combination thereof.

Furthermore, significant decreases in wave speed are observed upon removal of confining pressure, indicating the accumulation of sample damage. Decompression induced wave-speed decreases are greatest in experiments performed at high pressure (600~MPa) and low temperature (room temperature) in conjunction with the highest hardening rates. Microstructural observations from samples deformed at these conditions are characterised by open grain boundaries, suggesting that wave-speed decreases during decompression originate from the release of stored elastic strain. 

Electron microscopy shows that twin density is high, consistent with previous studies at similar conditions. Twin spacing is always at least one order of magnitude smaller than grain size. The spatial distributions of intragranular misorientation suggest that twin boundaries do not always act as significant barriers to dislocation motion and slip-transfer computations indicate that twins are statistically weaker barriers than grain boundaries. Therefore, grain size exerts a first-order control on strength and strain hardening, whereas the spacing of twin boundaries may exert a second-order control on these properties.

Taken together, our results show that semi-brittle flow in calcite is controlled by grain-size dependent processes that lead to significant hardening. This behaviour is qualitatively consistent with rheological models that include dislocation density and twin spacing as key state variables. The role of cracking in the decrease of strain hardening at low pressure and temperature requires the addition of a quantity describing crack density (that should include information on crack spacing, length, and orientation distribution) as a new state variable, to address fully the stress-strain behaviour of rocks in the semi-brittle regime. 

\section*{Open Research Section}
Processed experimental data (stress, strain and velocity change) is available from Zenodo (https://doi.org/10.5281/zenodo.7347236).

\acknowledgments
Extensive discussions with Erik Rybacki and Brian Evans, who shared some of their (then) unpublished data, helped to shape this work. Emmanuel David contributed to early technical developments on the Murrell apparatus. Technical support from John Bowles and Neil Hughes is greatly appreciated. Ian Jackson kindly provided the Wombeyan marble. Sarah Incel facilitated thin section preparation. Sheng Fan helped running SEM sessions. Discussions with Thomas Breithaupt, J\"org Renner and Chris Spiers contributed to our understanding of plastic deformation in calcite. This project has received funding from the European Research Council (ERC) under the European Union's Horizon 2020 research and innovation programme (grant agreement no 804685/“RockDEaF” to N.B.)  and from the UK Natural Environment Research Council (Grant Agreement No. NE/M016471/1 to N.B.). DW acknowledges support from a UK Research and Innovation Future Leaders Fellowship [grant number MR/V021788/1].

%
%

\bibliography{library}

%
%
%
%
%

\end{document}